\documentclass[aip,jcp,twocolumn]{revtex4}
\usepackage{graphicx}
\usepackage{longtable}
\usepackage{dcolumn}

\begin{document}
\newcommand{\cm}{cm$^{-1}$}
\newcommand{\B}{$B\,^1\Sigma^+_{\mathrm{u}}$}
\newcommand{\Bp}{$B'\,^1\Sigma^+_{\mathrm{u}}$}
\newcommand{\Bpp}{$B''\,^1\Sigma^+_{\mathrm{u}}$}
\newcommand{\C}{$C\,^1\Pi_{\mathrm{u}}$}
\newcommand{\Cplus}{$C\,^1\Pi_{\mathrm{u}}^+$}
\newcommand{\Cminus}{$C\,^1\Pi_{\mathrm{u}}^-$}
\newcommand{\D}{$D\,^1\Pi_{\mathrm{u}}$}
\newcommand{\Dplus}{$D\,^1\Pi_{\mathrm{u}}^+$}
\newcommand{\Dminus}{$D\,^1\Pi_{\mathrm{u}}^-$}
\newcommand{\Dp}{$D'\,^1\Pi_{\mathrm{u}}$}
\newcommand{\Dpp}{$D''\,^1\Pi_{\mathrm{u}}$}
\newcommand{\X}{$X\,^1\Sigma^+_{\mathrm{g}}$}
\newcommand{\dash}{\multicolumn{1}{c}{--}}

\title{VUV Fourier-transform absorption study of the Lyman and Werner bands in D$_2$}
\author{Arno de Lange}
\email{A.de.Lange@sron.nl}
\altaffiliation[Current address: ]{SRON - Netherlands Institute for Space Research, Sorbonnelaan 2, 3584 CA Utrecht, The Netherlands}
\affiliation{Institute for Lasers, Life and Biophotonics, VU University, De Boelelaan 1081, 1081 HV Amsterdam, The Netherlands}
\author{Gareth D. Dickenson}
\affiliation{Institute for Lasers, Life and Biophotonics, VU University, De Boelelaan 1081, 1081 HV Amsterdam, The Netherlands}
\author{Edcel J. Salumbides}
\affiliation{Institute for Lasers, Life and Biophotonics, VU University, De Boelelaan 1081, 1081 HV Amsterdam, The Netherlands}
\author{Wim Ubachs}
\affiliation{Institute for Lasers, Life and Biophotonics, VU University, De Boelelaan 1081, 1081 HV Amsterdam, The Netherlands}
\author{Nelson de Oliveira}
\affiliation{Synchrotron Soleil, Orme des Merisiers, St Aubin BP 48, 91192, GIF sur Yvette cedex, France}
\author{Denis Joyeux}
\affiliation{Synchrotron Soleil, Orme des Merisiers, St Aubin BP 48, 91192, GIF sur Yvette cedex, France}
\author{Laurent Nahon}
\affiliation{Synchrotron Soleil, Orme des Merisiers, St Aubin BP 48, 91192, GIF sur Yvette cedex, France}

\date{\today}
\begin{abstract}
An extensive survey of the D$_2$ absorption spectrum has been performed with the high-resolution VUV Fourier-transform spectrometer of the DESIRS beamline at the SOLEIL synchrotron.
The frequency range of $90\,000$--$119\,000$~\cm\ covers the full depth of the potential wells of the \B, \Bp, and \C\ electronic states up to the D(1s) + D($2\ell$) dissociation limit.
Improved level energies of rovibrational levels have been determined up to respectively $v=51$, $v=13$, and $v=20$.
Highest resolution is achieved by probing absorption in a molecular gas jet with slit geometry, as well as in a liquid helium cooled static gas cell, resulting in line widths of $\approx0.35$~\cm.
Extended calibration methods are employed to extract line positions of D$_2$ lines at absolute accuracies of $0.03$~\cm.
The \D\ and \Bpp\ electronic states correlate with the D(1s) + D($3\ell$) dissociation limit, but support a few vibrational levels below the second dissociation limit, respectively $v=0$--3 and $v=0$--1, and are also included in the presented study.
The complete set of resulting level energies is the most comprehensive and accurate data set for D$_2$.
The observations are compared with previous studies, both experimental and theoretical.
\end{abstract}
\pacs{pacs_numbers}

\maketitle

\section{Introduction}
Molecular hydrogen is the smallest neutral molecule and is as such a benchmark system for testing quantum mechanical calculations in molecules, starting from Born-Oppenheimer potentials, adiabatic and non-adiabatic corrections, leading to accurate predictions of level energies for all three natural isotopologues of hydrogen~\cite{Wolniewicz_1995}. Recently, also high-order relativistic and quantum-electrodynamic effects, i.e. molecular Lamb shifts were included in the calculations, although limited to the \X~ground state~\cite{Komasa_2011}. For D$_2$ these calculations were subjected to test and confirmed in a measurement of the dissociation energy of the ground state~\cite{Liu_2010} at an accuracy level of $< 0.001$ \cm. For the electronically excited states of $^1\Sigma_u^+$ symmetry~\cite{Staszewska_2002} and $^1\Pi_u$ symmetry~\cite{Wolniewicz_2003} \emph{ab initio} calculations have been performed, although at lower accuracy than for the ground state.

Due to the low nuclear masses in hydrogenic systems the validity of the Born-Oppenheimer approximation is only limited, less than in heavier molecules. Hence, isotopic effects are strong and the pronounced phenomena of mass-dependent adiabatic and non-adiabatic corrections can be well studied by comparing H$_2$, HD and D$_2$, where HD exhibits additional effects of breaking of the inversion symmetry~\cite{DeLange_2002}. For these reasons there is a continued interest in the investigation of the spectroscopy of hydrogen and its isotopomers, having started over a century ago by Lyman~\cite{Lyman_1906}. In particular the \B--\X~Lyman and \C--\X~Werner systems have attracted much attention, since these are the strongest, dipole-allowed, absorption systems originating from the \X\ electronic ground state. 

Spectroscopic studies on D$_2$ specifically bear relevance for the detailed investigation of thermonuclear fusion plasma reactors.
For example Hollmann~\emph{et al.} have detected extermely hot D$_2$ molecules in the DIII-D reactor from their spectroscopic signatures~\cite{Hollmann_2006}.
Similarly Pospieszczyk~\emph{et al.} investigated various hydrogen molecular isotopomers in the JET fusion reactor~\cite{Pospieszczyk_2007}.

The first vacuum ultraviolet absorption spectrum of D$_2$ was studied by Beutler~\emph{et al.}~\cite{Beutler_1935} in 1935 at relatively low resolution. From the 1960s, Herzberg and Monfils have studied its absorption spectrum over a wider range with a much higher accuracy, which led to the discovery of new electronic states (\Bpp, \Dp, and \Dpp) \cite{Herzberg_1960,Monfils_1965,Monfils_1968}.
In subsequent years Wilkinson~\cite{Wilkinson_1968}, Bredohl and Herzberg~\cite{Bredohl_1973}, Dabrowski and Herzberg~\cite{Dabrowski_1974}, Takezawa and Tanaka~\cite{Takezawa_1975}, and Larzilli\`ere~\emph{et al.}~\cite{Larzilliere_1980} have further extended the spectral investigations using classical spectrometers.

Later, after the development of nonlinear optical techniques, tunable extreme ultraviolet radiation from a laser-based source was used to yield improved accuracy in the spectroscopy of the D$_2$ Lyman and Werner bands~\cite{Hinnen_1994}. Over the years the accuracy has been further improved resulting in a highly accurate laser study by Roudjane~\emph{et al.}~\cite{Roudjane_2008}, focusing on a low number of bands, which may be used for calibration purposes of subsequent studies, including the present one. The most accurate comprehensive investigations of the D$_2$ spectrum were conducted by Abgrall~\emph{et al.}~\cite{Abgrall_1999} and Roudjane~\emph{et al.}~\cite{Roudjane_2006,Roudjane_2007}, both in emission and with spectrographs of 3~m and 10~m respectively.
Another extensive study is based on the emission data by Dieke's laboratory group, collected over 30 years starting in the early 1930s. This data was analyzed and published by Freund~\emph{et al.}~\cite{Freund_1985}.

Here we present a comprehensive absorption study of the D$_2$ spectrum, employing the high-resolution VUV Fourier-transform spectrometer at the SOLEIL synchrotron. All three electronic singlet states of \emph{ungerade} symmetry correlating with the D(1s) + D($2\ell$) dissociation limit are investigated; the \B, \Bp, and \C~states. 
Rovibrational levels have been observed over the full potential well depths for vibrational levels up to respectively $v=51$, $v=13$, and $v=20$, with an absolute accuracy of $0.03$~\cm. 
Some electronic states converging to the D(1s) + D($3\ell$) dissociation limit, \D\ and \Bpp, also exhibit rovibrational levels below the second dissociation limit, which are also listed for completeness. Predissociative resonances above the $n=2$ dissociation limit of D$_2$ have been published separately~\cite{Dickenson_2011}.

\section{Experimental setup}
The D$_2$ absorption spectra have been recorded in the gas phase at the synchrotron facility SOLEIL, where a vacuum ultraviolet (VUV) Fourier-Transform Spectrometer (FTS)~\cite{Oliveira_2009,Oliveira_2011} has been installed as a permanent instrument on the VUV undulator-based DESIRS beamline~\cite{Nahon_2012}. This FTS provides a high resolving power of $\approx 10^6$ over the entire instrumental wavelength range of $40$--$180$~nm covering the windowless regime of relevance for the present study. 
The undulator of the DESIRS beamline delivers broadband radiation with a bell-shaped spectrum, spanning $\approx 12\,000$~\cm, used as a continuum background feeding the FTS which central frequency is tuneable by changing the magnetic field of the undulator. The total frequency range investigated in the present study is $90\,000$--$119\,000$~\cm\ and overlapping spectra are recorded for covering this wide frequency range. 

Upstream of the FTS, the sample environment chamber is located, containing different types of gas-sample setups, upstream and downstream of which two similarly-sized holes ensure an efficient differential pumping with respect to the FTS chamber and the rest of the beamline. 
The FTS sample environment chamber is equipped with a free flow T-shaped gas cell containing the gas sample under quasi-static conditions.
The cell is either cooled down with liquid nitrogen (L-N$_2$) or with liquid helium (L-He) to reduce Doppler broadening. 
This absorption facility was also used in a previous investigation on the Lyman and Werner bands of the HD molecule~\cite{Ivanov_2010}.
In the present study a third type of measurement is performed in addition to the gas cell setup with two different coolants. 
The FTS is used to record absorption spectra from a D$_2$ molecular gas jet for the first time.

\begin{figure}[t]
\begin{center}
\includegraphics[width=0.9\columnwidth]{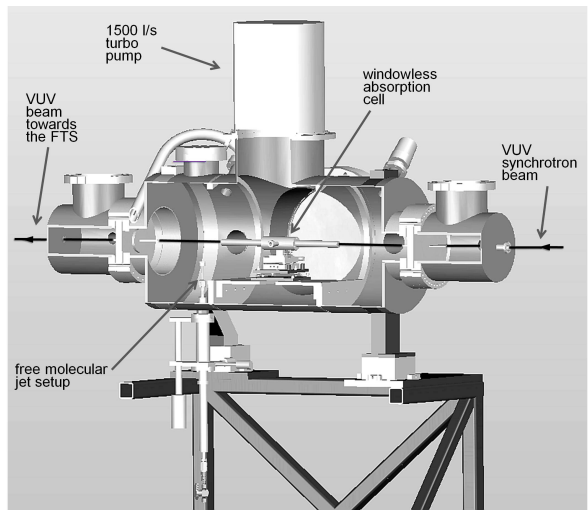}
\end{center}
\caption
{\label{fig_setup_jet}
  The FTS branch gas sample chamber on the DESIRS beamline at the SOLEIL synchrotron.
  The chamber includes a windowless absorption cell that can be cooled down thanks to a continuous flow of L-N$_2$ or L-He, plus, a free molecular jet setup.
  For clarity, the separate differentially-pumped jet expansion chamber and the setup for cooling the windowless cell are omitted from the picture.
  }
\end{figure}

The free molecular jet is located downstream of the windowless gas cell (see Fig.~\ref{fig_setup_jet} for experimental setup). The supersonic free expansion takes place in a separate chamber pumped continuously by a $500$~L/s turbo-molecular pump. The synchrotron beam passes through two holes in the expansion chamber that approximately fit the dimensions of the beam to limit the vacuum conductance. A nozzle slit shape ($1\,000\times5$~$\mu$m$^2$) is used, oriented so that the photon beam propagates along the slit length.
The backing pressure has been set such that saturation on lines of interest is avoided; due to pumping limitations it is not possible to exceed backing pressures beyond $6$~bar. It appears that the highest cold column density is observed when the photon beam crosses the molecular jet as close as possible to the nozzle position. Nevertheless, despite the two stages of differential pumping, background gas at room temperature can be seen on the absorption spectrum as a broad pedestal on which the narrow line is superimposed.

\begin{figure}[t]
\begin{center}
\includegraphics[width=0.9\columnwidth]{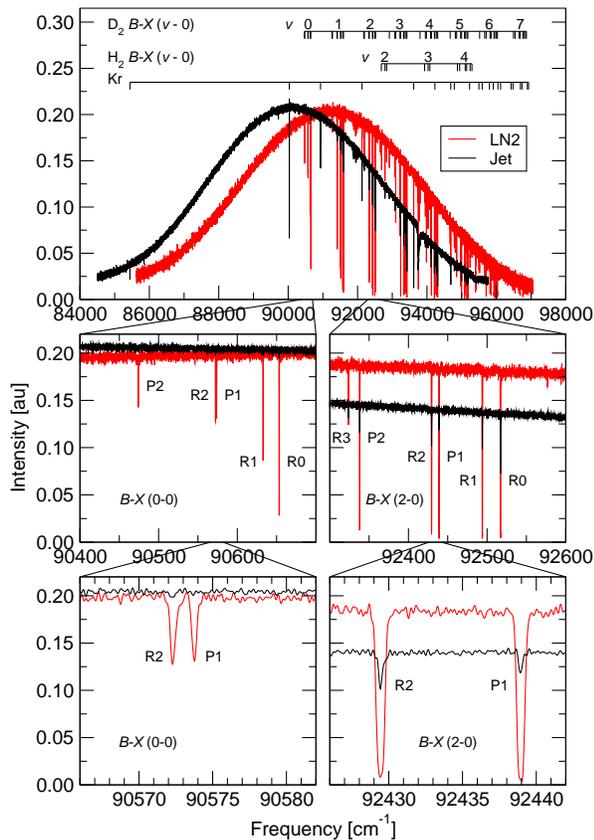}
\end{center}
\caption
{\label{fig_spectrum}
  D$_2$ absorption spectra recorded in a jet (black) and a cell cooled by liquid nitrogen (red).
  From the static gas cell setup many more lines are discernible.
  The lower panels zoom into two bands: \B--\X~($0$--$0$) (left) and \B--\X~($2$--$0$) (right). Note that a series of
  lines exciting Rydberg states of the Kr atom are included in some of the spectra; these lines originate from the gas filter used for eliminating harmonic radiation produced by the undulator.
  For further details see text.
  }
\end{figure}

Figure~\ref{fig_spectrum} provides a view on the typical FT-spectral recordings, with two slightly shifted bell-shaped undulator profiles shown in the top panel, and two stages of zooming to show details of the individual absorption lines of D$_2$. The black curves illustrate recordings with the molecular jet and the red curves illustrate measurements employing the L-N$_2$-cooled quasi-static gas cell. The Figure shows that in the L-N$_2$ cooled cell configuration many more lines are discernable than in the jet configuration. However, the lines exhibit narrower profiles with the jet.

Under the three different experimental conditions, \emph{i.e.} the L-N$_2$ and L-He cooled cell and the jet, different line widths are observed. These widths relate mainly to the resulting Doppler width, but also depend on the optical density at which the experiments are carried out.
The data set covers respectively 326 lines for L-He cooled, 472 lines for L-N$_2$ cooled, and 284 lines for the jet configurations. These data pertain to the frequency range up to the second dissociation limit ($119\,030$~\cm), to exclude lines possibly broadened by predissocation. For all three cases the line width distribution is not normal and exhibits a shoulder towards higher widths; this is most pronounced for the L-N$_2$ case due to saturation effects. Discarding the saturated lines, the average means of the line widths, when fitting with a single Gaussian, are $0.35$~\cm\ for the jet, $0.37$~\cm\ for the L-He cooled cell, and $0.48$~\cm\ for the L-N$_2$ cooled cell. The widths are a convolution of contributions of the instrument profile (a sinc-function of $0.16$~\cm\ width related to the settings and travel arm of the FT-instrument), the effective Doppler width resulting from the inhomogeneously distributed outward diffusing gas in the cooled T-shaped cell, and a small additional broadening due to possible beam pointing instability during the FT-recordings.
In the line width analysis above, the contribution of the background gas at room temperature is disregarded.
Its effect is a broad pedestal on which the narrow(er) absorption line is superimposed and when not accounted for it effectively broadens the line. This effect is observed in particular in case of the jet.

In the recorded spectra, some H$_2$ lines are observed as well. 
The widths of these lines are $0.87$~\cm, which is twice as broad as the unsaturated D$_2$ lines.
This is due to a larger Doppler width, which stems mainly from the fact that the H$_2$ resides in the background gas at room temperature, but also from the lower mass of H$_2$.

\section{Frequency calibration}
\label{Calibration}
The Fourier-transform spectra exhibit an internal frequency calibration derived directly from the interferogram sampling intervals and determined for each spectrum by an interferometric measurement using a stabilized Helium-Neon laser~\cite{Oliveira_2009,Oliveira_2011,Ivanov_2010}.
Due to small alignment offsets of the Helium-Neon laser and the VUV beam relative directions, that may vary from run to run, the absolute calibration also varies for different runs, and can be improved upon by anchoring the spectra to several accurately known D$_2$ lines in addition to a few Xe and Kr lines that occur in the spectra, finding their origin in the gas filter used for attenuating the harmonics at short wavelengths produced in the undulator.
Some 39 lines belonging to the \B$\,(v'=9-11)$--\X$\,(v''=0)$ and \C$\,(v'=0)$--\X$\,(v''=0)$ systems, previously measured using an extreme ultraviolet laser instrument by Roudjane~\emph{et al.}~\cite{Roudjane_2008} at an accuracy of $0.006$~\cm, are used for this purpose. 
The Kr and Xe lines are taken from Refs.~\cite{Brandi_2002,Brandi_2001}.
The calibration procedure is repeated for each scan in order to remove any possible variation. 
In practice, the correction may vary slightly over long periods of time. 
After correction, the spread in the differences between the presently observed FT-line frequencies and the laser-based frequencies of Ref.~\cite{Roudjane_2008} is $0.02$~\cm, and this is taken as the statistical error for the present data set.

However, these calibration lines fall within $96\,000$--$100\,000$~\cm, a range that is only covered by the scans at low frequencies.
Therefore, an extrapolation towards higher frequencies is required for the absolute calibration of the remaining scans.
The absolute frequency scale of subsequent overlapping scans is adapted by overlaying a large number of lines ($> 30$), yielding sufficient statistics to achieve a relative uncertainty in the frequency scale of $\approx 0.003$~\cm\ between two adjacent scans. 
Towards higher frequencies, this procedure is applied multiple times, increasing the uncertainty with every step.
The largest uncertainty pertains thus to the scan with the highest frequencies and amounts to $0.009$~\cm.
Based on this value, the systematic error is conservatively estimated to be $0.01$~\cm\ for all scans. 
The uncertainty in the absolute frequencies for all lines is thus estimated at $0.03$~\cm; $0.02$~\cm\ statistical plus $0.01$~\cm\ systematic error.

\begin{figure}[t]
\begin{center}
\includegraphics[width=0.9\columnwidth]{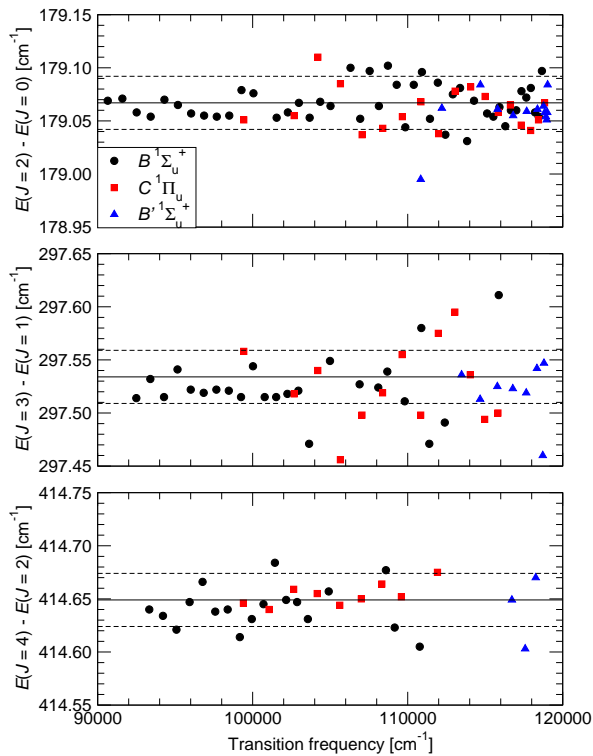}
\end{center}
\caption
{\label{fig_comb_diff}
  The combination differences of various transition pairs from a wide range of vibrational levels in the \B\ (black circles), \C\ (red squares), and \Bp\ (blue triangles) electronic states of D$_2$.
  The following pairs are depicted; $P(2)-R(0)$ (upper panel), $P(3)-R(1)$ (central panel), and $P(4)-R(2)$ (lower panel).
  The solid lines correspond to the theoretical values of the combination differences, taken from Ref.~\cite{Komasa_2011}. 
  The dashed lines indicate the standard deviation of $0.025$~\cm\ of all the differences.
  The transition frequencies refer to the $R$ branch.
}
\end{figure}

In Fig.~\ref{fig_comb_diff} electronic ground state combination differences are plotted for the line combinations $P(2)-R(0)$, $P(3)-R(1)$ and $P(4)-R(2)$, as measured in transitions to all vibrational levels in the \B, \C, and \Bp\ states.
Note that blended lines are excluded from this plot. 
The solid lines in these plots refer to the most accurate theoretical combination differences from Ref.~\cite{Komasa_2011}, yielding $\Delta_{20}=179.067$ \cm, $\Delta_{31}=297.534$ \cm, and $\Delta_{42}=414.649$ \cm.
The observed combination differences agree very well with these theoretical values and the standard deviation for all differences is $0.025$~\cm.
This is in good agreement with the estimated statistical uncertainty of $0.02$~\cm\ for a single line.
In fact, it is even slightly lower than expected as the uncertainty in the combination differences from two transitions is $\sqrt{2}\times\,0.02 = 0.03$~\cm.
Note that only the statistical error is taken into account since the corresponding $P$ and $R$ transitions are sufficiently close in frequency to cancel any systematic errors in the combination differences. 

In addition, the scatter in line frequencies of strong D$_2$ lines, \emph{i.e.} lines with a $\mathrm{S/N} > 6$, that are observed in multiple runs is also $0.02$~\cm, validating the statistical uncertainty estimate.
However, for weaker lines (with S/N $<$ 6) this scatter increases to $0.05$~\cm, and hence the total uncertainty for these lines is estimated at $0.06$~\cm.

To test the validity of the systematic error estimate of $0.01$~\cm, the observed H$_2$ lines are compared with the highly accurate data by Bailly~\emph{et al.}~\cite{Bailly_2010}.
The standard deviation of these differences is $0.05$~\cm, and is therefore more than twice as large as the observed scattering of $0.02$~\cm\ in the D$_2$ lines.
This can be explained by the fact that the observed H$_2$ lines are also more than twice as broad as the unsaturated D$_2$ lines.
Because of this the H$_2$ lines are not used for the absolute calibration.
The average of the $47$ differences between the H$_2$ lines in the present study and the data by Bailly~\emph{et al.} is $0.005(7)$~\cm.
Thus the two data sets agree within this uncertainty and also the estimated systematic uncertainty of $0.01$~\cm\ is consistent with this comparison.

\section{Results}

The dipole-allowed absorption spectrum of D$_2$, in the range up to the $n=2$ dissociation limit, where narrow unpredissociated resonances are found, is recorded in absorption. Observed transition frequencies over the full depth of the potential wells of the \B, \Bp, and \C\ electronic states, converging to the $n=2$ limit of D$_2$ are presented. The vibrational levels $v=0$--$51$ have been observed in the \B\ state, $v=0$--$13$ in the \Bp\ state and $v=0$--$20$ in the \C\ state. In addition the vibrational levels of the unpredissociated \D\ and \Bpp\ states that lie below the second dissociation limit are presented as well; $v=0$--$3$ for \D\ and $v=0$--$1$ for \Bpp, respectively. Extensive lists of all observed transition frequencies are given in the supplementary material data depository of the American Institute of Physics~\cite{SUPP}.

Many of the measured lines have been observed before, albeit at lower accuracy. There are, however, a few levels probed for the first time. In case of the \B\ state, Freund~\emph{et al.}~\cite{Freund_1985} present level energies for the vibrational levels $v=48$ and $v=50$, but not $v=49$. This might have been caused by blending of lines; in our study the $R(1)$ and $R(2)$ transitions are blended, while the $R(0)$ and $P(1)$ transitions are well-resolved. Of the last observed $v=51$ vibrational level Dabrowski and Herzberg found the $R(0)$ transition~\cite{Dabrowski_1974}, whereas in the present study the $R(0)$, $R(1)$, and $P(1)$ transitions have been observed. The observed vibrational levels $v=0$--13 in the \Bp\ electronic state have all been observed before by Dabrowski and Herzberg~\cite{Dabrowski_1974}, Freund~\emph{et al.}~\cite{Freund_1985}, and Abgrall~\emph{et al.}~\cite{Abgrall_1999}. In case of the \C\ electronic state, all vibrational levels $v=0$--20 were observed before by Dabrowski and Herzberg~\cite{Dabrowski_1974}, but Freund~\emph{et al.}~\cite{Freund_1985} and Abgrall~\emph{et al.}~\cite{Abgrall_1999} observed only vibrational levels $v=0$--18.

The information content of the measured transition frequencies is condensed to values for the level energies.
For those levels probed by multiple transitions, the uncertainty in the level energy is conservatively taken as the highest accuracy of these transitions, rather than an average.
The highly accurate excitation energies of rotational levels in the \X\ ground state are taken from Ref.~\cite{Komasa_2011}. All resulting level energies for the five excited states of singlet and $ungerade$ symmetry below the second dissociation limit of D$_2$ as probed in this study are listed in Tables~\ref{table_line_list_B}--\ref{table_line_list_Bpp}: Table~\ref{table_line_list_B} lists the data for the \B\ state, Table~\ref{table_line_list_Bp} for the \Bp\ state, Table~\ref{table_line_list_C} for the \C\ state, Table~\ref{table_line_list_D} for the \D\ state, and Table~\ref{table_line_list_Bpp} for the \Bpp\ state. 
In order to present the Tables updated to the most accurate values, the highly accurate laser data of Roudjane~\emph{et al.}~\cite{Roudjane_2008} are included in Tables~\ref{table_line_list_B} and \ref{table_line_list_C}.

\begin{table*}
\caption
{\label{table_line_list_B}
  Observed level energies for the \B\ state in \cm, relative to the \X~$(v=0,J=0)$ level.
  The uncertainties in the last digit are indicated in superscript and level energies that have been derived from blended lines, are listed with $^b$. The highly accurate laser date by Roudjane~\emph{et al.}~\cite{Roudjane_2008} are included in this Table and marked with $^l$.
}
\begin{ruledtabular}
\begin{tabular}{rddddddd}
$v$ &    J=0       &    J=1       &    J=2       &    J=3       &    J=4       &    J=5       &    J=6    \\
\hline
  0 &  90633.47^3 &  90653.18^3 &  90692.52^3 &  90751.31^3 &  90829.14^6 & \dash        & \dash \\
  1 &  91575.79^3 &  91594.78^3 &  91632.69^3 &  91689.32^3 &  91764.42^3 &  91857.61^6 & \dash \\
  2 &  92498.66^3 &  92517.03^3 &  92553.68^3 &  92608.45^3 &  92681.08^3 &  92771.33^3 & \dash \\
  3 &  93403.30^3 &  93421.10^3 &  93456.62^3 &  93509.71^3 &  93580.14^3 &  93667.65^3 & \dash \\
  4 &  94290.39^3 &  94307.67^3 &  94342.14^3 &  94393.70^3 &  94462.13^3 &  94547.11^3 & \dash \\
  5 &  95160.37^3 &  95177.16^3 &  95210.68^3 &  95260.82^3 &  95327.35^3 &  95410.03^3 & \dash \\
  6 &  96013.54^3 &  96029.88^3 &  96062.51^3 &  96111.30^3 &  96176.05^3 &  96256.52^3 & \dash \\
  7 &  96850.19^3 &  96866.12^3 &  96897.88^3 &  96945.41^3 &  97008.46^3 &  97086.85^3 & \dash \\
  8 &  97670.47^3 &  97685.98^3 &  97716.93^3 &  97763.23^3 &  97824.67^3 &  97901.09^3 & \dash \\
  9 &  98474.533^{7,l}&  98489.648^{6,l}&  98519.825^{6,l}&  98564.953^{6,l}&  98624.871^{7,l}&  98699.377^{6,l}& \dash \\
 10 &  99262.597^{6,l}&  99277.320^{6,l}&  99306.726^{7,l}&  99350.723^{6,l}&  99409.151^{6,l}&  99481.813^{7,l}&  99568.46^6 \\
 11 & 100034.805^{7,l}& 100049.180^{6,l}& 100077.890^{6,l}& 100120.825^{6,l}& 100177.850^{6,l}& 100248.771^{6,l}& \dash \\
 12 & 100791.34^3 & 100805.30^3 & 100833.31^3 & 100875.18^3 & 100930.78^3 & 100999.94^3 & 101082.46^6 \\
 13 & 101532.37^3 & 101546.06^3 & 101573.41^3 & 101614.35^3 & 101668.70^3 & 101736.41^3 & \dash \\
 14 & 102258.08^3 & 102271.41^3 & 102298.04^3 & 102337.91^3 & 102390.87^3 & 102456.77^3 & \dash \\
 15 & 102968.63^3 & 102981.69^3 & 103007.79^3 & 103046.89^3 & 103098.92^3 & 103164.24^3 & 103229.34^6 \\
 16 & 103664.22^3 & 103676.90^3 & 103702.28^3 & 103740.28^3 & 103790.78^3 & 103853.58^3 & \dash \\
 17 & 104344.98^3 & 104357.52^3 & 104382.58^3 & 104420.42^3 & 104475.47^3 & 104526.83^3 & \dash \\
 18 & 105011.18^3 & 105023.31^3 & 105047.54^3 & 105083.77^3 & 105131.91^3 & \dash       & \dash \\
 19 & 105662.92^3 & 105676.38^3 & 105695.86^3 & 105732.17^3 & 105779.47^3 & 105838.11^3 & \dash \\
 20 & 106300.54^3 & 106312.10^3 & 106335.20^3 & 106369.80^3 & 106415.77^3 & \dash       & \dash \\
 21 & 106924.02^3 & 106935.15^3 & 106957.46^3 & 106990.99^3 & 107035.67^b & 107091.36^6 & \dash \\
 22 & 107533.68^3 & 107544.75^3 & 107566.77^3 & 107599.89^3 & 107643.83^3 & \dash       & \dash \\
 23 & 108129.68^3 & 108140.33^3 & 108161.69^3 & 108193.69^3 & 108236.34^3 & 108289.45^3 & \dash \\
 24 & 108712.16^3 & 108722.71^3 & 108743.81^3 & 108775.50^3 & 108817.76^6 & \dash       & \dash \\
 25 & 109281.28^3 & 109291.47^3 & 109311.89^3 & 109342.45^3 & 109383.11^3 & 109433.81^6 & \dash \\
 26 & 109837.24^3 & 109847.38^3 & 109867.67^3 & 109898.28^3 & \dash       & \dash       & \dash \\
 27 & 110380.18^3 & 110389.92^3 & 110409.40^3 & 110438.56^3 & \dash       & \dash       & \dash \\
 28 & 110910.25^3 & 110920.20^3 & 110940.85^3 & 110955.32^3 & 111000.72^3 & 111049.15^6 & \dash \\
 29 & 111427.59^3 & 111436.87^3 & 111455.40^6 & 111483.28^3 & 111520.25^3 & \dash       & \dash \\
 30 & 111932.28^3 & 111940.80^3 & 111958.39^3 & 111985.19^3 & 112021.03^6 & \dash       & \dash \\
 31 & 112424.43^3 & 112433.30^3 & 112451.01^3 & 112477.57^3 & 112512.91^6 & \dash       & \dash \\
 32 & 112904.16^3 & 112912.58^3 & 112929.54^3 & 112955.03^3 & 112989.05^6 & \dash       & \dash \\
 33 & 113371.54^b & 113379.86^3 & 113396.76^3 & 113422.07^3 & 113455.91^6 & \dash       & \dash \\
 34 & 113826.28^3 & 113834.38^3 & 113850.53^3 & 113874.79^3 & \dash       & \dash       & \dash \\
 35 & 114268.74^3 & 114276.76^3 & 114292.88^3 & 114317.09^3 & \dash       & \dash       & \dash \\
 36 & 114698.64^3 & 114706.32^3 & 114721.66^3 & 114744.65^3 & 114775.33^6 & \dash       & \dash \\
 37 & 115115.91^3 & 115123.57^3 & 115138.96^3 & 115162.37^3 & \dash       & \dash       & \dash \\
 38 & 115520.27^3 & 115527.51^3 & 115542.00^3 & 115563.67^3 & \dash       & \dash       & \dash \\
 39 & 115911.40^3 & 115918.74^3 & 115933.71^3 & 115967.24^b & 115984.78^b & \dash       & \dash \\
 40 & 116288.85^3 & 116295.63^3 & 116309.18^3 & 116329.49^3 & 116356.43^6 & \dash       & \dash \\
 41 & 116651.94^3 & 116659.07^3 & 116678.24^b & 116685.29^3 & 116713.45^b & \dash       & \dash \\
 42 & 116999.78^3 & 117006.05^3 & 117018.59^3 & 117037.35^3 & \dash       & \dash       & \dash \\
 43 & 117331.22^3 & 117338.29^3 & 117343.51^3 & 117363.91^3 & 117388.22^6 & \dash       & \dash \\
 44 & 117644.59^3 & 117650.23^3 & 117661.57^3 & 117678.47^3 & \dash       & \dash       & \dash \\
 45 & 117937.59^3 & 117943.80^3 & 117948.20^3 & 117966.45^3 & \dash       & \dash       & \dash \\
 46 & 118207.17^3 & 118212.03^3 & 118222.21^3 & 118236.29^3 & \dash       & \dash       & \dash \\
 47 & 118448.94^3 & 118453.81^3 & 118465.33^3 & 118472.24^3 & \dash       & \dash       & \dash \\
 48 & 118656.86^3 & 118660.55^3 & 118668.01^3 & 118679.07^3 & \dash       & \dash       & \dash \\
 49 & 118822.43^3 & 118824.72^3 & 118830.72^b & 118839.59^b & \dash       & \dash       & \dash \\
 50 & 118934.57^6 & 118936.51^3 & 118940.29^3 & 118945.82^6 & \dash       & \dash       & \dash \\
 51 & 118988.95^6 & 118989.87^3 & 118991.74^b & \dash       & \dash       & \dash       & \dash \\
\end{tabular}
\end{ruledtabular}
\end{table*}
\begin{table*}
\caption
{\label{table_line_list_Bp}
  Observed level energies for the \Bp\ state in \cm, relative to the \X~$(v=0,J=0)$ level.
  The uncertainties in the last digit are indicated in superscript and level energies that have been derived from blended lines, are listed with $^b$.
}
\begin{ruledtabular}
\begin{tabular}{rdddddd}
$v$ &    J=0       &    J=1       &    J=2       &    J=3       &    J=4       &    J=5    \\
\hline
  0 & \dash       & 110841.89^3 & 110894.39^3 & 110972.71^3 & \dash       & \dash    \\
  1 & 112180.87^3 & 112205.74^3 & 112255.40^3 & 112329.68^3 & 112428.29^3 & 112550.89^6 \\
  2 & 113467.04^3 & 113490.86^3 & 113538.45^3 & 113609.70^3 & 113704.43^3 & \dash    \\
  3 & 114669.40^3 & 114690.74^3 & 114733.63^3 & 114798.35^3 & 114884.89^b & 114992.98^3 \\
  4 & 115779.82^3 & 115800.42^3 & 115841.55^3 & 115903.08^3 & 115984.78^b & 116086.36^3 \\
  5 & 116784.48^3 & 116803.71^3 & 116842.08^3 & 116899.49^3 & 116975.70^b & 117070.52^3 \\
  6 & 117659.28^3 & 117675.39^3 & 117707.81^3 & 117756.75^3 & 117821.99^3 & 117903.16^3 \\
  7 & 118357.31^3 & 118371.08^3 & 118398.42^3 & 118438.98^3 & 118492.18^3 & 118557.30^6 \\
  8 & 118754.82^3 & 118761.30^3 & 118773.70^6 & 118791.03^3 & \dash       & \dash    \\
  9 & 118838.52^3 & 118842.94^3 & 118852.22^3 & 118863.67^3 & 118881.06^3 & \dash    \\
 10 & 118913.11^3 & 118916.30^3 & 118922.61^3 & 118931.99^3 & 118944.24^6 & \dash    \\
 11 & 118966.66^3 & 118968.97^3 & 118973.56^3 & 118980.30^3 & \dash       & \dash    \\
 12 & 119003.47^3 & 119005.18^3 & 119008.49^3 & 119012.96^3 & \dash       & \dash    \\
 13 & 119027.27^3 & 119028.14^3 & 119029.62^3 & \dash       & \dash       & \dash    \\
\end{tabular}
\end{ruledtabular}
\end{table*}
\begin{table*}
\caption
{\label{table_line_list_C}
  Observed level energies for the \C\ state in \cm, relative to the \X~$(v=0,J=0)$ level.
  The uncertainties in the last digit are indicated in superscript and level energies that have been derived from blended lines, are listed with $^b$. The highly accurate laser date by Roudjane~\emph{et al.}~\cite{Roudjane_2008} are included in this Table and marked with $^l$.
}
\begin{ruledtabular}
\begin{tabular}{rdddddd}
$v$ &    J=1       &    J=2       &    J=3       &    J=4       &    J=5       &    J=6    \\
\hline
\\
& \multicolumn{6}{c}{\Cplus} \\
\\
 0  &  99424.957^{6,l}&  99486.942^{6,l}&  99579.584^{6,l}&  99702.495^{6,l}&  99855.174^{7,l}& \dash    \\
 1  & 101085.41^b & 101145.17^b & 101234.54^3 & 101353.05^3 & 101500.09^b & 101675.10^b \\
 2  & 102677.96^3 & 102735.57^3 & 102821.63^3 & 102935.59^3 & 103076.56^3 & \dash    \\
 3  & 104203.85^3 & 104259.09^3 & 104341.32^3 & 104446.30^3 & 104592.26^3 & \dash    \\
 4  & 105662.55^3 & 105720.18^3 & 105798.95^3 & 105904.43^3 & 106035.49^3 & \dash    \\
 5  & 107059.38^3 & 107110.86^3 & 107187.59^3 & 107289.11^b & 107414.91^3 & 107565.57^b \\
 6  & 108389.67^3 & 108438.85^3 & 108512.20^3 & 108609.17^3 & 108728.91^3 & \dash    \\
 7  & 109655.36^3 & 109702.18^3 & 109771.80^3 & 109862.93^3 & 109965.30^3 & \dash    \\
 8  & 110855.96^3 & 110899.27^3 & 110980.32^3 & 111061.62^3 & 111170.21^3 & \dash    \\
 9  & 111992.28^3 & 112035.48^3 & 112099.47^3 & 112183.90^3 & 112288.36^3 & \dash    \\
10  & 113060.72^3 & 113101.37^3 & 113161.78^3 & 113241.53^3 & \dash       & \dash    \\
11  & 114061.24^3 & 114099.35^3 & 114156.04^3 & 114230.56^3 & \dash       & \dash    \\
12  & 114991.35^3 & 115026.78^3 & 115079.31^3 & 115146.75^3 & \dash       & \dash    \\
13  & 115847.97^3 & 115880.38^3 & 115926.71^3 & \dash       & \dash       & \dash    \\
14  & 116626.94^3 & 116654.85^3 & 116709.53^3 & 116767.32^6 & \dash       & \dash    \\
15  & 117322.89^3 & 117357.30^3 & 117395.93^3 & 117449.55^6 & \dash       & \dash    \\
16  & 117929.64^3 & 117960.26^3 & 117994.06^3 & 118041.33^3 & \dash       & \dash    \\
17  & 118437.01^3 & 118454.90^3 & 118491.68^3 & \dash       & \dash       & \dash    \\
18  & 118831.41^3 & 118847.08^3 & 118872.50^6 & \dash       & \dash       & \dash    \\
19  & 119085.91^3 & 119095.78^3 & 119111.03^b & \dash       & \dash       & \dash    \\
\\
& \multicolumn{6}{c}{\Cminus} \\
\\
 0  &  99424.672^{6,l}&  99486.108^{7,l}&  99577.958^{6,l}&  99699.867^{6,l}& \dash       & \dash    \\
 1  & 101085.05^3 & 101144.25^3 & 101232.72^3 & 101350.16^3 & 101495.96^3 & \dash    \\
 2  & 102677.63^3 & 102734.61^3 & 102819.80^3 & 102932.83^3 & 103073.31^3 & \dash    \\
 3  & 104203.55^3 & 104258.34^3 & 104340.28^3 & 104449.00^3 & \dash       & \dash    \\
 4  & 105663.72^3 & 105716.40^3 & 105795.13^3 & 105899.59^3 & 106029.24^6 & \dash    \\
 5  & 107058.84^3 & 107110.86^b & 107184.93^3 & 107285.13^3 & 107409.59^3 & \dash    \\
 6  & 108389.25^3 & 108437.70^3 & 108510.03^3 & 108606.02^3 & 108725.28^3 & \dash    \\
 7  & 109655.05^3 & 109701.35^3 & 109770.53^3 & 109862.33^3 & 109976.27^6 & \dash    \\
 8  & 110855.99^3 & 110900.14^3 & 110966.14^3 & 111053.67^3 & \dash       & \dash    \\
 9  & 111991.39^3 & 112033.41^3 & 112096.17^3 & 112179.39^3 & 112282.89^b & \dash    \\
10  & 113060.24^3 & 113100.05^3 & 113159.51^3 & 113238.38^3 & \dash       & \dash    \\
11  & 114060.91^3 & 114098.46^3 & 114154.52^3 & \dash       & \dash       & \dash    \\
12  & 114991.15^3 & 115026.35^3 & 115078.82^3 & \dash       & \dash       & \dash    \\
13  & 115847.95^b & 115880.67^3 & 115929.49^3 & \dash       & \dash       & \dash    \\
14  & 116627.27^3 & 116657.35^3 & 116702.22^3 & \dash       & \dash       & \dash    \\
15  & 117323.76^3 & 117350.98^3 & 117391.54^3 & 117445.00^3 & \dash       & \dash    \\
16  & 117930.35^3 & 117954.39^3 & 117990.23^6 & \dash       & \dash       & \dash    \\
17  & 118437.34^3 & 118457.76^3 & 118486.41^b & \dash       & \dash       & \dash    \\
18  & 118830.72^b & 118846.90^3 & 118870.76^6 & \dash       & \dash       & \dash    \\
19  & 119085.82^3 & \dash       & \dash       & \dash       & \dash       & \dash    \\
20  & 119157.00^6 & \dash       & \dash       & \dash       & \dash       & \dash    \\
\end{tabular}
\end{ruledtabular}
\end{table*}
\begin{table*}
\caption
{\label{table_line_list_D}
  Observed level energies for the \D\ state in \cm, relative to the \X~$(v=0,J=0)$ level.
  The uncertainties in the last digit are indicated in superscript.
}
\begin{ruledtabular}
\begin{tabular}{rddddd}
$v$ &    J=1       &    J=2       &    J=3       &    J=4       &    J=5    \\
\hline
\\
& \multicolumn{5}{c}{\Dplus} \\
\\
  0 & 113222.94^3 & 113283.38^3 & 113373.68^3 & 113493.18^6 & \dash    \\
  1 & 114825.06^3 & 114885.13^3 & 114974.54^3 & 115092.29^3 & 115239.24^3 \\
  2 & 116359.51^3 & 116415.90^3 & 116500.02^6 & 116611.33^3 & 116748.96^6 \\
  3 & 117831.40^3 & 117886.79^3 & 117968.89^3 & 118076.91^3 & 118209.98^3 \\
\\
& \multicolumn{5}{c}{\Dminus} \\
\\
  0 & 113222.42^3 & 113281.91^3 & 113370.85^3 & 113488.88^6 & \dash    \\
  1 & 114823.44^3 & 114880.73^3 & 114966.37^3 & \dash       & \dash    \\
  2 & 116358.82^3 & 116413.95^3 & 116496.35^3 & 116605.73^3 & \dash    \\
  3 & 117830.03^3 & 117883.05^3 & 117962.28^3 & 118067.44^3 & \dash    \\
\end{tabular}
\end{ruledtabular}
\end{table*}
\begin{table*}
\caption
{\label{table_line_list_Bpp}
  Observed level energies for the \Bpp\ state in \cm, relative to the \X~$(v=0,J=0)$ level.
  The uncertainties in the last digit are indicated in superscript.
}
\begin{ruledtabular}
\begin{tabular}{rddddd}
$v$ &    J=0       &    J=1       &    J=2       &    J=3       &    J=4    \\
\hline
  0 & 117197.17^3 & 117224.14^3 & 117278.03^3 & 117358.68^3 & 117465.87^6 \\
  1 & 118688.02^3 & 118714.41^3 & 118767.09^3 & 118845.98^3 & 118950.81^3 \\
\end{tabular}
\end{ruledtabular}
\end{table*}

\section{Discussion}

The present set of level energies comprises the most accurate comprehensive data set for the five electronic states (\B, \Bp, \Bpp, \C\ and \D\ states) of singlet and \emph{ungerade} symmetry supporting bound levels below the $n=2$ dissociation limit in D$_2$. It is of interest to compare these accurate determinations of experimental level energies with those from previous studies and to those predicted by theory. It is noted that a comparison with the accurate laser data by Roudjane~\emph{et al.}~\cite{Roudjane_2008} is made implicitly since those data are used for calibration of the presently recorded spectra.

\begin{figure}[t]
\begin{center}
\includegraphics[width=0.95\columnwidth]{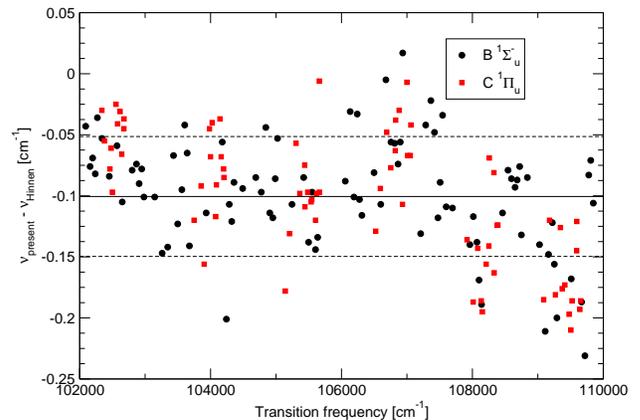}
\end{center}
\caption
{\label{fig_hinnen_differences}
  Deviations between observed transitions in the present study and the observed transition frequencies by Hinnen~\emph{et al.}~\cite{Hinnen_1994}.
  Black circles refer to transitions to the \B\ state, whereas the red squares pertain to the \C\ state.
}
\end{figure}

First a comparison is made with the laser data by Hinnen~\emph{et al.}~\cite{Hinnen_1994}, a comprehensive data set with claimed accuracies of $0.03$--$0.08$~\cm.
It is noted that the XUV-laser system in this study is based on a PDL (pulsed dye laser) system as opposed to the PDA (pulsed dye amplifier) system used by Roudjane~\emph{et al.}~\cite{Roudjane_2008}.
The instrument width with the PDL is much larger than with the PDA, and subsequently leads to a lower accuracy. 
For technical details see also Ref.~\cite{Ubachs_1997}.
The comparison between the present study and the data by Hinnen is shown in Fig.~\ref{fig_hinnen_differences}. 
The solid line is the averaged difference and amounts to $-0.10$~\cm, with the dashed lines the $1\sigma = 0.05$~\cm\ spread in the differences.
The values by Hinnen~\emph{et al.} are thus systematically higher than in the present study.
A similar difference ($-0.07$~\cm) has also been observed in the case of H$_2$ as pointed out by Philip~\emph{et al.}~\cite{Philip_2004}.
It is therefore believed that this systematic offset is due to the PDL-laser setup and the calibration procedure used in Ref.~\cite{Hinnen_1994}.

\begin{figure}[t]
\begin{center}
\includegraphics[width=0.95\columnwidth]{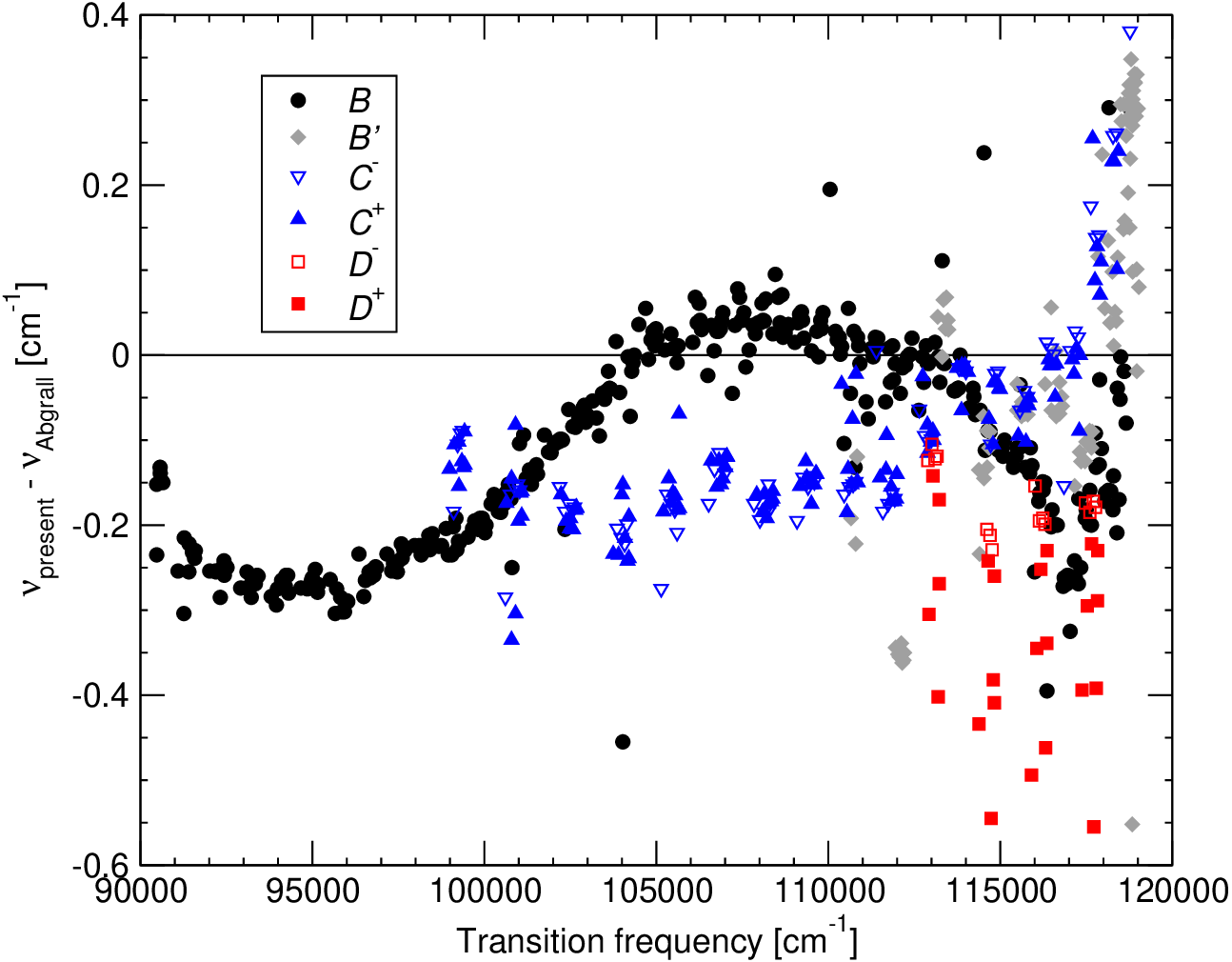}
\end{center}
\caption
{\label{fig_Abgrall_differences}
  Deviations between observed level energies in the present study and the calculated values by Abgrall~\emph{et al.}~\cite{Abgrall_1999}.
  The black circles refer to the \B\ electronic state, the grey diamonds to \Bp, the blue triangles to \C, and the red squares to \D.
  The open triangles and squares pertain correspond respectively to the \Cminus\ and \Dminus\ electronic states, whereas the filled shapes to \Cplus\ and \Dplus.
  The solid black line at $0$~\cm\ is to guide the eye, and clearly shows that the values in the present study are mostly lower than those by Abgrall~\emph{et al.}~\cite{Abgrall_1999}.
}
\end{figure}

In the work of Abgrall~\emph{et al.}~\cite{Abgrall_1999} a calculation is performed including non-adiabatic interaction effects in a four-state analysis for the (e)-parity levels (\B, \Bp, \Cplus\ and \Dplus\ states); the (f)-parity levels can be treated separately in a two-state analysis involving \Cminus\ and \Dminus\ states. These calculations were performed, based on the Born-Oppenheimer potential curves by Dressler and Wolniewicz~\cite{Dressler_1986} and the \emph{ab initio} calculations of the non-adiabatic couplings~\cite{Senn_1988,Wolniewicz_1992}. In a study of the emission spectrum of D$_2$ the existing potentials were semi-empirically optimized by fitting to line intensities and line positions in the spectrum, resulting in a slightly deviating potential energy curve~\cite{Abgrall_1999}.
Experimental line positions were taken from the analysis by Freund~\emph{et al.}~\cite{Freund_1985}, who analysed the extensive emission data set from Dieke's laboratory group.
In Fig.~\ref{fig_Abgrall_differences} the differences between the experimental level energies, as determined in the current study, with respect to the calculations by Abgrall~\emph{et al.}~\cite{Abgrall_1999}, are shown for the \B, \Bp, \C\ and \D\ electronic states.
The systematic deviation of $0.2$~\cm\ between the present experimental results and the calculations in the lower frequency range  ($< 100\,000$~\cm) is ascribed to an offset in the experimental values in the emission study; the theoretical values in Ref.~\cite{Abgrall_1999} are adapted to the experimental ones via a fit of the potential. An absolute calibration uncertainty of $0.2$~\cm\ is not surprising for a classical spectrometer study. The deviation of about $0.15$~\cm\ in the frequency range $105\,000$--$110\,000$~\cm\ between levels pertaining to \B\ and \C\ states is more surprising, since these points derive from the same part of the spectrum and relate to relative errors. The scatter in the data points for the \C\ state is most likely to be ascribed to non-adiabatic interactions with the \B\ state, modeled only to a certain extent. In the frequency range $> 113\,000$~\cm\ the scatter in the data points becomes much larger (even as large as $\pm 0.2$~\cm); here the modeling of non-adiabatic interactions between the four states of (e) symmetry is the limiting factor. This conclusion is supported by the fact that the levels pertaining to the \Dminus\ state (f-symmetry) show much less spread per vibrational level than for the \Dplus\ state; indeed the \Dminus\ state only interacts with the \Cminus\ state, while the \Dplus\ state is part of a four state interaction. In fact, in the region $>117\,000$~\cm\ the \Bpp\ state perturbs the (e)-levels and is not accounted for in the theoretical model.
Over the whole frequency range, the \B\ state shows hardly any scatter in the data points, which can be interpreted that this state is only weakly perturbed by other states, apart from a few incidental local interactions.
There are some outliers in the differences between the data set by Abgrall~\emph{et al.}~\cite{Abgrall_1999} and the present study.
However, almost all either pertain to blended lines in the present study, or show a similar difference between the fitted values by Abgrall~\emph{et al.}~\cite{Abgrall_1999} and the measured transition frequencies as given by Freund~\emph{et al.}~\cite{Freund_1985}. 
This indicates that the modeled spectra do not fully capture all level interactions.

\begin{figure}[t]
\begin{center}
\includegraphics[width=0.95\columnwidth]{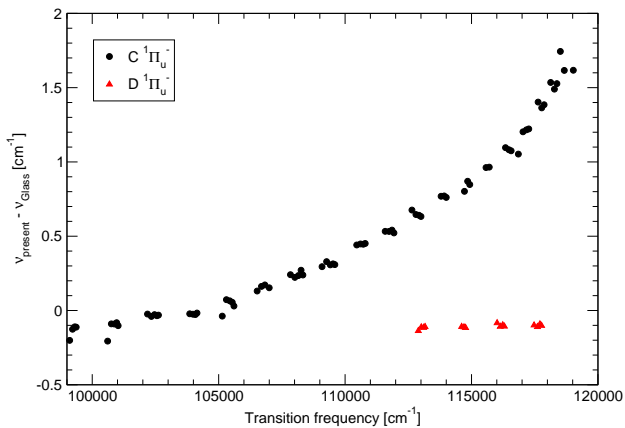}
\end{center}
\caption
{\label{fig_Cm_Dm_mqdt}
  Deviations between observed level energies in the present study and those of the MQDT calculations by Glass-Maujean~\emph{et al.}~\cite{Glass_2011}.
  The black circles correspond to $Q$ transitions to the \Cminus\ electronic state and the red triangles to \Dminus.
}
\end{figure}

In Fig.~\ref{fig_Cm_Dm_mqdt} a comparison is made between the presently observed data and those obtained from another theoretical framework, the multi-channel quantum defect (MQDT) formalism. The MQDT-formalism was developed by Jungen and Atabek~\cite{Jungen_1977} to describe the level structure of \C\ and \D\ states of the hydrogen molecule. This framework has recently been further refined by Glass-Maujean~\emph{et al.} and compared with accurate data on emission in hydrogen and deuterium, focusing on \Cminus\ and \Dminus\ levels~\cite{Glass_2011}. 
The MQDT-calculations are in good agreement with the present measurements, and can be seen to be accurate to within 1.5~\cm\ for both the \Cminus\ and \Dminus\ electronic states.
For low vibrational levels, occurring deeply in the potential wells, the differences are even smaller, and are only 0.1--0.2~\cm.

\section{Conclusion}

High resolution spectra of the D$_2$ molecule have been recorded with the VUV Fourier-transform spectrometer at the DESIRS beamline at the SOLEIL synchrotron.
For the first time a slit jet geometry was combined with the VUV-FTS to achieve a spectral resolution of $0.35$~\cm, while spectra of similar quality were obtained employing a liquid-He cooled quasi-static gas cell.
The present study delivers the most comprehensive and accurate data set for the \B, \Bp, and \C\ electronic states in D$_2$ covering the entire depth of the potential wells below the $n=2$ dissociation limit.
In addition the sharp unpredissociated levels of the \D\ and \Bpp\ states are included.
Line positions are determined, and level energies extracted, at an absolute accuracy of $0.03$~\cm, which corresponds to a fractional uncertainty of $3 \times 10^{-7}$.

\section*{Acknowledgments}

The authors would like to thank Dr. Alan Heays for fruitful discussions on the data analysis. The Netherlands Foundation for Fundamental Research of Matter (FOM) is gratefully acknowledged for financial support. 
We are grateful to the general staff of SOLEIL for running the facility and in particular to Jean-Fran{\c c}ois Gil for his technical help with the sample environment chamber.


\begin{thebibliography}{40}
\expandafter\ifx\csname natexlab\endcsname\relax\def\natexlab#1{#1}\fi
\expandafter\ifx\csname bibnamefont\endcsname\relax
  \def\bibnamefont#1{#1}\fi
\expandafter\ifx\csname bibfnamefont\endcsname\relax
  \def\bibfnamefont#1{#1}\fi
\expandafter\ifx\csname citenamefont\endcsname\relax
  \def\citenamefont#1{#1}\fi
\expandafter\ifx\csname url\endcsname\relax
  \def\url#1{\texttt{#1}}\fi
\expandafter\ifx\csname urlprefix\endcsname\relax\def\urlprefix{URL }\fi
\providecommand{\bibinfo}[2]{#2}
\providecommand{\eprint}[2][]{\url{#2}}

\bibitem[{\citenamefont{Wolniewicz}(1995)}]{Wolniewicz_1995}
\bibinfo{author}{\bibfnamefont{L.}~\bibnamefont{Wolniewicz}},
  \bibinfo{journal}{J.\ Chem.\ Phys.} \textbf{\bibinfo{volume}{103}},
  \bibinfo{pages}{1792} (\bibinfo{year}{1995}).

\bibitem[{\citenamefont{Komasa et~al.}(2011)\citenamefont{Komasa,
  Piszczatowski, {\L}ach, Przybytek, Jeziorski, and Pachucki}}]{Komasa_2011}
\bibinfo{author}{\bibfnamefont{J.}~\bibnamefont{Komasa}},
  \bibinfo{author}{\bibfnamefont{K.}~\bibnamefont{Piszczatowski}},
  \bibinfo{author}{\bibfnamefont{G.}~\bibnamefont{{\L}ach}},
  \bibinfo{author}{\bibfnamefont{M.}~\bibnamefont{Przybytek}},
  \bibinfo{author}{\bibfnamefont{B.}~\bibnamefont{Jeziorski}},
  \bibnamefont{and} \bibinfo{author}{\bibfnamefont{K.}~\bibnamefont{Pachucki}},
  \bibinfo{journal}{J.\ Chem.\ Theory\ Comput.} \textbf{\bibinfo{volume}{7}},
  \bibinfo{pages}{3105} (\bibinfo{year}{2011}).

\bibitem[{\citenamefont{Liu et~al.}(2010)\citenamefont{Liu, Sprecher, Jungen,
  Ubachs, and Merkt}}]{Liu_2010}
\bibinfo{author}{\bibfnamefont{J.}~\bibnamefont{Liu}},
  \bibinfo{author}{\bibfnamefont{D.}~\bibnamefont{Sprecher}},
  \bibinfo{author}{\bibfnamefont{C.}~\bibnamefont{Jungen}},
  \bibinfo{author}{\bibfnamefont{W.}~\bibnamefont{Ubachs}}, \bibnamefont{and}
  \bibinfo{author}{\bibfnamefont{F.}~\bibnamefont{Merkt}},
  \bibinfo{journal}{J.\ Chem.\ Phys.} \textbf{\bibinfo{volume}{132}},
  \bibinfo{pages}{154301} (\bibinfo{year}{2010}).

\bibitem[{\citenamefont{Staszewska and Wolniewicz}(2002)}]{Staszewska_2002}
\bibinfo{author}{\bibfnamefont{G.}~\bibnamefont{Staszewska}} \bibnamefont{and}
  \bibinfo{author}{\bibfnamefont{L.}~\bibnamefont{Wolniewicz}},
  \bibinfo{journal}{J.\ Mol.\ Spectr.} \textbf{\bibinfo{volume}{212}},
  \bibinfo{pages}{208} (\bibinfo{year}{2002}).

\bibitem[{\citenamefont{Wolniewicz and Staszewska}(2003)}]{Wolniewicz_2003}
\bibinfo{author}{\bibfnamefont{L.}~\bibnamefont{Wolniewicz}} \bibnamefont{and}
  \bibinfo{author}{\bibfnamefont{G.}~\bibnamefont{Staszewska}},
  \bibinfo{journal}{J.\ Mol.\ Spectr.} \textbf{\bibinfo{volume}{220}},
  \bibinfo{pages}{45} (\bibinfo{year}{2003}).

\bibitem[{\citenamefont{de~Lange et~al.}(2002)\citenamefont{de~Lange, Reinhold,
  and Ubachs}}]{DeLange_2002}
\bibinfo{author}{\bibfnamefont{A.}~\bibnamefont{de~Lange}},
  \bibinfo{author}{\bibfnamefont{E.}~\bibnamefont{Reinhold}}, \bibnamefont{and}
  \bibinfo{author}{\bibfnamefont{W.}~\bibnamefont{Ubachs}},
  \bibinfo{journal}{Int.\ Rev.\ Phys.\ Chem.} \textbf{\bibinfo{volume}{21}},
  \bibinfo{pages}{257} (\bibinfo{year}{2002}).

\bibitem[{\citenamefont{Lyman}(1906)}]{Lyman_1906}
\bibinfo{author}{\bibfnamefont{T.}~\bibnamefont{Lyman}},
  \bibinfo{journal}{Astroph.\ J.} \textbf{\bibinfo{volume}{23}},
  \bibinfo{pages}{181} (\bibinfo{year}{1906}).

\bibitem[{\citenamefont{Hollmann et~al.}(2006)\citenamefont{Hollmann,
  Brezinsek, Brooks, Groth, McLean, Pigarov, and Rudakov}}]{Hollmann_2006}
\bibinfo{author}{\bibfnamefont{E.~M.} \bibnamefont{Hollmann}},
  \bibinfo{author}{\bibfnamefont{S.}~\bibnamefont{Brezinsek}},
  \bibinfo{author}{\bibfnamefont{N.~H.} \bibnamefont{Brooks}},
  \bibinfo{author}{\bibfnamefont{M.}~\bibnamefont{Groth}},
  \bibinfo{author}{\bibfnamefont{A.~G.} \bibnamefont{McLean}},
  \bibinfo{author}{\bibfnamefont{A.~Y.} \bibnamefont{Pigarov}},
  \bibnamefont{and} \bibinfo{author}{\bibfnamefont{D.~L.}
  \bibnamefont{Rudakov}}, \bibinfo{journal}{Plasma\ Phys.\ Control.\ Fusion}
  \textbf{\bibinfo{volume}{48}}, \bibinfo{pages}{1165} (\bibinfo{year}{2006}).

\bibitem[{\citenamefont{Pospieszczyk et~al.}(2007)\citenamefont{Pospieszczyk,
  Brezinsek, Meigs, Mertens, Sergienko, and Stamp}}]{Pospieszczyk_2007}
\bibinfo{author}{\bibfnamefont{A.}~\bibnamefont{Pospieszczyk}},
  \bibinfo{author}{\bibfnamefont{S.}~\bibnamefont{Brezinsek}},
  \bibinfo{author}{\bibfnamefont{A.}~\bibnamefont{Meigs}},
  \bibinfo{author}{\bibfnamefont{P.}~\bibnamefont{Mertens}},
  \bibinfo{author}{\bibfnamefont{G.}~\bibnamefont{Sergienko}},
  \bibnamefont{and} \bibinfo{author}{\bibfnamefont{M.}~\bibnamefont{Stamp}},
  \bibinfo{journal}{J.\ Nucl.\ Mater.} \textbf{\bibinfo{volume}{363}},
  \bibinfo{pages}{811} (\bibinfo{year}{2007}).

\bibitem[{\citenamefont{Beutler et~al.}(1935)\citenamefont{Beutler, Deubner,
  and J{\"u}nger}}]{Beutler_1935}
\bibinfo{author}{\bibfnamefont{H.}~\bibnamefont{Beutler}},
  \bibinfo{author}{\bibfnamefont{A.}~\bibnamefont{Deubner}}, \bibnamefont{and}
  \bibinfo{author}{\bibfnamefont{H.~O.} \bibnamefont{J{\"u}nger}},
  \bibinfo{journal}{Z.\ Phys.} \textbf{\bibinfo{volume}{98}},
  \bibinfo{pages}{181} (\bibinfo{year}{1935}).

\bibitem[{\citenamefont{Herzberg and Monfils}(1960)}]{Herzberg_1960}
\bibinfo{author}{\bibfnamefont{G.}~\bibnamefont{Herzberg}} \bibnamefont{and}
  \bibinfo{author}{\bibfnamefont{A.}~\bibnamefont{Monfils}},
  \bibinfo{journal}{J.\ Mol.\ Spectros.} \textbf{\bibinfo{volume}{5}},
  \bibinfo{pages}{482} (\bibinfo{year}{1960}).

\bibitem[{\citenamefont{Monfils}(1965)}]{Monfils_1965}
\bibinfo{author}{\bibfnamefont{A.}~\bibnamefont{Monfils}},
  \bibinfo{journal}{J.\ Mol.\ Spectros.} \textbf{\bibinfo{volume}{15}},
  \bibinfo{pages}{265} (\bibinfo{year}{1965}).

\bibitem[{\citenamefont{Monfils}(1968)}]{Monfils_1968}
\bibinfo{author}{\bibfnamefont{A.}~\bibnamefont{Monfils}},
  \bibinfo{journal}{J.\ Mol.\ Spectros.} \textbf{\bibinfo{volume}{25}},
  \bibinfo{pages}{513} (\bibinfo{year}{1968}).

\bibitem[{\citenamefont{Wilkinson}(1968)}]{Wilkinson_1968}
\bibinfo{author}{\bibfnamefont{P.~G.} \bibnamefont{Wilkinson}},
  \bibinfo{journal}{Can.\ J.\ Phys.} \textbf{\bibinfo{volume}{46}},
  \bibinfo{pages}{1225} (\bibinfo{year}{1968}).

\bibitem[{\citenamefont{Bredohl and Herzberg}(1973)}]{Bredohl_1973}
\bibinfo{author}{\bibfnamefont{H.}~\bibnamefont{Bredohl}} \bibnamefont{and}
  \bibinfo{author}{\bibfnamefont{G.}~\bibnamefont{Herzberg}},
  \bibinfo{journal}{Can.\ J.\ Phys.} \textbf{\bibinfo{volume}{51}},
  \bibinfo{pages}{867} (\bibinfo{year}{1973}).

\bibitem[{\citenamefont{Dabrowski and Herzberg}(1974)}]{Dabrowski_1974}
\bibinfo{author}{\bibfnamefont{I.}~\bibnamefont{Dabrowski}} \bibnamefont{and}
  \bibinfo{author}{\bibfnamefont{G.}~\bibnamefont{Herzberg}},
  \bibinfo{journal}{Can.\ J.\ Phys.} \textbf{\bibinfo{volume}{52}},
  \bibinfo{pages}{1110} (\bibinfo{year}{1974}).

\bibitem[{\citenamefont{Takezawa and Tanaka}(1975)}]{Takezawa_1975}
\bibinfo{author}{\bibfnamefont{S.}~\bibnamefont{Takezawa}} \bibnamefont{and}
  \bibinfo{author}{\bibfnamefont{Y.}~\bibnamefont{Tanaka}},
  \bibinfo{journal}{J.\ Mol.\ Spectros.} \textbf{\bibinfo{volume}{54}},
  \bibinfo{pages}{379} (\bibinfo{year}{1975}).

\bibitem[{\citenamefont{Larzilli\`ere et~al.}(1980)\citenamefont{Larzilli\`ere,
  Launay, and Roncin}}]{Larzilliere_1980}
\bibinfo{author}{\bibfnamefont{M.}~\bibnamefont{Larzilli\`ere}},
  \bibinfo{author}{\bibfnamefont{F.}~\bibnamefont{Launay}}, \bibnamefont{and}
  \bibinfo{author}{\bibfnamefont{J.-Y.} \bibnamefont{Roncin}},
  \bibinfo{journal}{J.\ Phys.\ (Paris)} \textbf{\bibinfo{volume}{41}},
  \bibinfo{pages}{1431} (\bibinfo{year}{1980}).

\bibitem[{\citenamefont{Hinnen et~al.}(1994)\citenamefont{Hinnen, Hogervorst,
  Stolte, and Ubachs}}]{Hinnen_1994}
\bibinfo{author}{\bibfnamefont{P.~C.} \bibnamefont{Hinnen}},
  \bibinfo{author}{\bibfnamefont{W.}~\bibnamefont{Hogervorst}},
  \bibinfo{author}{\bibfnamefont{S.}~\bibnamefont{Stolte}}, \bibnamefont{and}
  \bibinfo{author}{\bibfnamefont{W.}~\bibnamefont{Ubachs}},
  \bibinfo{journal}{Can.\ J.\ Phys.} \textbf{\bibinfo{volume}{72}},
  \bibinfo{pages}{1032} (\bibinfo{year}{1994}).

\bibitem[{\citenamefont{Roudjane et~al.}(2008)\citenamefont{Roudjane, Ivanov,
  Vieitez, de~Lange, Tchang-Brillet, and Ubachs}}]{Roudjane_2008}
\bibinfo{author}{\bibfnamefont{M.}~\bibnamefont{Roudjane}},
  \bibinfo{author}{\bibfnamefont{T.~I.} \bibnamefont{Ivanov}},
  \bibinfo{author}{\bibfnamefont{M.~O.} \bibnamefont{Vieitez}},
  \bibinfo{author}{\bibfnamefont{C.~A.} \bibnamefont{de~Lange}},
  \bibinfo{author}{\bibfnamefont{W.-{\"U}.~L.} \bibnamefont{Tchang-Brillet}},
  \bibnamefont{and} \bibinfo{author}{\bibfnamefont{W.}~\bibnamefont{Ubachs}},
  \bibinfo{journal}{Mol.\ Phys.} \textbf{\bibinfo{volume}{106}},
  \bibinfo{pages}{1193} (\bibinfo{year}{2008}).

\bibitem[{\citenamefont{Abgrall et~al.}(1999)\citenamefont{Abgrall, Roueff,
  Liu, Shemansky, and James}}]{Abgrall_1999}
\bibinfo{author}{\bibfnamefont{H.}~\bibnamefont{Abgrall}},
  \bibinfo{author}{\bibfnamefont{E.}~\bibnamefont{Roueff}},
  \bibinfo{author}{\bibfnamefont{X.}~\bibnamefont{Liu}},
  \bibinfo{author}{\bibfnamefont{D.~E.} \bibnamefont{Shemansky}},
  \bibnamefont{and} \bibinfo{author}{\bibfnamefont{G.~K.} \bibnamefont{James}},
  \bibinfo{journal}{J.\ Phys.\ B} \textbf{\bibinfo{volume}{32}},
  \bibinfo{pages}{3813} (\bibinfo{year}{1999}).

\bibitem[{\citenamefont{Roudjane et~al.}(2006)\citenamefont{Roudjane, Launay,
  and Tchang-Brillet}}]{Roudjane_2006}
\bibinfo{author}{\bibfnamefont{M.}~\bibnamefont{Roudjane}},
  \bibinfo{author}{\bibfnamefont{F.}~\bibnamefont{Launay}}, \bibnamefont{and}
  \bibinfo{author}{\bibfnamefont{W.-{\"U}.~L.} \bibnamefont{Tchang-Brillet}},
  \bibinfo{journal}{J.\ Chem.\ Phys.} \textbf{\bibinfo{volume}{125}},
  \bibinfo{pages}{214305} (\bibinfo{year}{2006}).

\bibitem[{\citenamefont{Roudjane et~al.}(2007)\citenamefont{Roudjane,
  Tchang-Brillet, and Launay}}]{Roudjane_2007}
\bibinfo{author}{\bibfnamefont{M.}~\bibnamefont{Roudjane}},
  \bibinfo{author}{\bibfnamefont{W.-{\"U}.~L.} \bibnamefont{Tchang-Brillet}},
  \bibnamefont{and} \bibinfo{author}{\bibfnamefont{F.}~\bibnamefont{Launay}},
  \bibinfo{journal}{J.\ Chem.\ Phys.} \textbf{\bibinfo{volume}{127}},
  \bibinfo{pages}{054307} (\bibinfo{year}{2007}).

\bibitem[{\citenamefont{Freund et~al.}(1985)\citenamefont{Freund, Schiavone,
  and Crosswhite}}]{Freund_1985}
\bibinfo{author}{\bibfnamefont{R.~S.} \bibnamefont{Freund}},
  \bibinfo{author}{\bibfnamefont{J.~A.} \bibnamefont{Schiavone}},
  \bibnamefont{and} \bibinfo{author}{\bibfnamefont{H.~M.}
  \bibnamefont{Crosswhite}}, \bibinfo{journal}{J.\ Phys.\ Chem.\ Ref.\ Data}
  \textbf{\bibinfo{volume}{14}}, \bibinfo{pages}{235} (\bibinfo{year}{1985}).

\bibitem[{\citenamefont{Dickenson et~al.}(2011)\citenamefont{Dickenson, Ivanov,
  Ubachs, Roudjane, de~Oliveira, Joyeux, Nahon, Tchang-Brillet, Glass-Maujean,
  Schmoranzer et~al.}}]{Dickenson_2011}
\bibinfo{author}{\bibfnamefont{G.~D.} \bibnamefont{Dickenson}},
  \bibinfo{author}{\bibfnamefont{T.~I.} \bibnamefont{Ivanov}},
  \bibinfo{author}{\bibfnamefont{W.}~\bibnamefont{Ubachs}},
  \bibinfo{author}{\bibfnamefont{M.}~\bibnamefont{Roudjane}},
  \bibinfo{author}{\bibfnamefont{N.}~\bibnamefont{de~Oliveira}},
  \bibinfo{author}{\bibfnamefont{D.}~\bibnamefont{Joyeux}},
  \bibinfo{author}{\bibfnamefont{L.}~\bibnamefont{Nahon}},
  \bibinfo{author}{\bibfnamefont{W.-{\"U}.~L.} \bibnamefont{Tchang-Brillet}},
  \bibinfo{author}{\bibfnamefont{M.}~\bibnamefont{Glass-Maujean}},
  \bibinfo{author}{\bibfnamefont{H.}~\bibnamefont{Schmoranzer}},
  \bibnamefont{et~al.}, \bibinfo{journal}{Mol.\ Phys.}
  \textbf{\bibinfo{volume}{109}}, \bibinfo{pages}{2693} (\bibinfo{year}{2011}).

\bibitem[{\citenamefont{de~Oliveira et~al.}(2009)\citenamefont{de~Oliveira,
  Joyeux, Phalippou, Rodier, Polack, Vervloet, and Nahon}}]{Oliveira_2009}
\bibinfo{author}{\bibfnamefont{N.}~\bibnamefont{de~Oliveira}},
  \bibinfo{author}{\bibfnamefont{D.}~\bibnamefont{Joyeux}},
  \bibinfo{author}{\bibfnamefont{D.}~\bibnamefont{Phalippou}},
  \bibinfo{author}{\bibfnamefont{J.~C.} \bibnamefont{Rodier}},
  \bibinfo{author}{\bibfnamefont{F.}~\bibnamefont{Polack}},
  \bibinfo{author}{\bibfnamefont{M.}~\bibnamefont{Vervloet}}, \bibnamefont{and}
  \bibinfo{author}{\bibfnamefont{L.}~\bibnamefont{Nahon}},
  \bibinfo{journal}{Rev.\ Sci.\ Instrum.} \textbf{\bibinfo{volume}{80}},
  \bibinfo{pages}{043101} (\bibinfo{year}{2009}).

\bibitem[{\citenamefont{de~Oliveira et~al.}(2011)\citenamefont{de~Oliveira,
  Roudjane, Joyeux, Phalippou, Rodier, and Nahon}}]{Oliveira_2011}
\bibinfo{author}{\bibfnamefont{N.}~\bibnamefont{de~Oliveira}},
  \bibinfo{author}{\bibfnamefont{M.}~\bibnamefont{Roudjane}},
  \bibinfo{author}{\bibfnamefont{D.}~\bibnamefont{Joyeux}},
  \bibinfo{author}{\bibfnamefont{D.}~\bibnamefont{Phalippou}},
  \bibinfo{author}{\bibfnamefont{J.~C.} \bibnamefont{Rodier}},
  \bibnamefont{and} \bibinfo{author}{\bibfnamefont{L.}~\bibnamefont{Nahon}},
  \bibinfo{journal}{Nat.\ Photon.} \textbf{\bibinfo{volume}{5}},
  \bibinfo{pages}{149} (\bibinfo{year}{2011}).

\bibitem[{\citenamefont{Nahon et~al.}(in press)\citenamefont{Nahon,
  de~Oliveira, Garcia, Gil, Pilette, Marcouill\'{e}, Lagarde, and
  Polack}}]{Nahon_2012}
\bibinfo{author}{\bibfnamefont{L.}~\bibnamefont{Nahon}},
  \bibinfo{author}{\bibfnamefont{N.}~\bibnamefont{de~Oliveira}},
  \bibinfo{author}{\bibfnamefont{G.}~\bibnamefont{Garcia}},
  \bibinfo{author}{\bibfnamefont{J.~F.} \bibnamefont{Gil}},
  \bibinfo{author}{\bibfnamefont{B.}~\bibnamefont{Pilette}},
  \bibinfo{author}{\bibfnamefont{O.}~\bibnamefont{Marcouill\'{e}}},
  \bibinfo{author}{\bibfnamefont{B.}~\bibnamefont{Lagarde}}, \bibnamefont{and}
  \bibinfo{author}{\bibfnamefont{F.}~\bibnamefont{Polack}},
  \bibinfo{journal}{J.\ Synchrotron Radiat.}  (\bibinfo{year}{in press}).

\bibitem[{\citenamefont{Ivanov et~al.}(2010)\citenamefont{Ivanov, Dickenson,
  de~Oliveira, Roudjane, Joyeux, Nahon, Tchang-Brillet, and
  Ubachs}}]{Ivanov_2010}
\bibinfo{author}{\bibfnamefont{T.~I.} \bibnamefont{Ivanov}},
  \bibinfo{author}{\bibfnamefont{G.~D.} \bibnamefont{Dickenson}},
  \bibinfo{author}{\bibfnamefont{N.}~\bibnamefont{de~Oliveira}},
  \bibinfo{author}{\bibfnamefont{M.}~\bibnamefont{Roudjane}},
  \bibinfo{author}{\bibfnamefont{D.}~\bibnamefont{Joyeux}},
  \bibinfo{author}{\bibfnamefont{L.}~\bibnamefont{Nahon}},
  \bibinfo{author}{\bibfnamefont{W.-{\"U}.~L.} \bibnamefont{Tchang-Brillet}},
  \bibnamefont{and} \bibinfo{author}{\bibfnamefont{W.}~\bibnamefont{Ubachs}},
  \bibinfo{journal}{Mol.\ Phys.} \textbf{\bibinfo{volume}{108}},
  \bibinfo{pages}{771} (\bibinfo{year}{2010}).

\bibitem[{\citenamefont{Brandi et~al.}(2002)\citenamefont{Brandi, Hogervorst,
  and Ubachs}}]{Brandi_2002}
\bibinfo{author}{\bibfnamefont{F.}~\bibnamefont{Brandi}},
  \bibinfo{author}{\bibfnamefont{W.}~\bibnamefont{Hogervorst}},
  \bibnamefont{and} \bibinfo{author}{\bibfnamefont{W.}~\bibnamefont{Ubachs}},
  \bibinfo{journal}{J.\ Phys.\ B} \textbf{\bibinfo{volume}{35}},
  \bibinfo{pages}{1071} (\bibinfo{year}{2002}).

\bibitem[{\citenamefont{Brandi et~al.}(2001)\citenamefont{Brandi, Velchev,
  Hogervorst, and Ubachs}}]{Brandi_2001}
\bibinfo{author}{\bibfnamefont{F.}~\bibnamefont{Brandi}},
  \bibinfo{author}{\bibfnamefont{I.}~\bibnamefont{Velchev}},
  \bibinfo{author}{\bibfnamefont{W.}~\bibnamefont{Hogervorst}},
  \bibnamefont{and} \bibinfo{author}{\bibfnamefont{W.}~\bibnamefont{Ubachs}},
  \bibinfo{journal}{Phys.\ Rev.\ A} \textbf{\bibinfo{volume}{64}},
  \bibinfo{pages}{032505} (\bibinfo{year}{2001}).

\bibitem[{\citenamefont{Bailly et~al.}(2010)\citenamefont{Bailly, Salumbides,
  Vervloet, and Ubachs}}]{Bailly_2010}
\bibinfo{author}{\bibfnamefont{D.}~\bibnamefont{Bailly}},
  \bibinfo{author}{\bibfnamefont{E.~J.} \bibnamefont{Salumbides}},
  \bibinfo{author}{\bibfnamefont{M.}~\bibnamefont{Vervloet}}, \bibnamefont{and}
  \bibinfo{author}{\bibfnamefont{W.}~\bibnamefont{Ubachs}},
  \bibinfo{journal}{Mol.\ Phys.} \textbf{\bibinfo{volume}{108}},
  \bibinfo{pages}{827} (\bibinfo{year}{2010}).

\bibitem[{SUP()}]{SUPP}
\bibinfo{howpublished}{See supplementary material at http://doi.xxx.xxx for all
  measured transition frequencies.}

\bibitem[{\citenamefont{Ubachs et~al.}(1997)\citenamefont{Ubachs, Eikema,
  Hogervorst, and Cacciani}}]{Ubachs_1997}
\bibinfo{author}{\bibfnamefont{W.}~\bibnamefont{Ubachs}},
  \bibinfo{author}{\bibfnamefont{K.~S.~E.} \bibnamefont{Eikema}},
  \bibinfo{author}{\bibfnamefont{W.}~\bibnamefont{Hogervorst}},
  \bibnamefont{and} \bibinfo{author}{\bibfnamefont{P.~C.}
  \bibnamefont{Cacciani}}, \bibinfo{journal}{J.\ Opt.\ Soc.\ Am.\ B}
  \textbf{\bibinfo{volume}{14}}, \bibinfo{pages}{2469} (\bibinfo{year}{1997}).

\bibitem[{\citenamefont{Philip et~al.}(2004)\citenamefont{Philip, Sprengers,
  Pielage, de~Lange, Ubachs, and Reinhold}}]{Philip_2004}
\bibinfo{author}{\bibfnamefont{J.}~\bibnamefont{Philip}},
  \bibinfo{author}{\bibfnamefont{J.~P.} \bibnamefont{Sprengers}},
  \bibinfo{author}{\bibfnamefont{T.}~\bibnamefont{Pielage}},
  \bibinfo{author}{\bibfnamefont{C.~A.} \bibnamefont{de~Lange}},
  \bibinfo{author}{\bibfnamefont{W.}~\bibnamefont{Ubachs}}, \bibnamefont{and}
  \bibinfo{author}{\bibfnamefont{E.}~\bibnamefont{Reinhold}},
  \bibinfo{journal}{Can.\ J.\ Chem.} \textbf{\bibinfo{volume}{82}},
  \bibinfo{pages}{713} (\bibinfo{year}{2004}).

\bibitem[{\citenamefont{Dressler and Wolniewicz}(1986)}]{Dressler_1986}
\bibinfo{author}{\bibfnamefont{K.}~\bibnamefont{Dressler}} \bibnamefont{and}
  \bibinfo{author}{\bibfnamefont{L.}~\bibnamefont{Wolniewicz}},
  \bibinfo{journal}{J.\ Chem.\ Phys.} \textbf{\bibinfo{volume}{85}},
  \bibinfo{pages}{2821} (\bibinfo{year}{1986}).

\bibitem[{\citenamefont{Senn et~al.}(1988)\citenamefont{Senn, Quadrelli, and
  Dressler}}]{Senn_1988}
\bibinfo{author}{\bibfnamefont{P.}~\bibnamefont{Senn}},
  \bibinfo{author}{\bibfnamefont{P.}~\bibnamefont{Quadrelli}},
  \bibnamefont{and} \bibinfo{author}{\bibfnamefont{K.}~\bibnamefont{Dressler}},
  \bibinfo{journal}{J.\ Chem.\ Phys.} \textbf{\bibinfo{volume}{89}},
  \bibinfo{pages}{7401} (\bibinfo{year}{1988}).

\bibitem[{\citenamefont{Wolniewicz and Dressler}(1992)}]{Wolniewicz_1992}
\bibinfo{author}{\bibfnamefont{L.}~\bibnamefont{Wolniewicz}} \bibnamefont{and}
  \bibinfo{author}{\bibfnamefont{K.}~\bibnamefont{Dressler}},
  \bibinfo{journal}{J.\ Chem.\ Phys.} \textbf{\bibinfo{volume}{96}},
  \bibinfo{pages}{6053} (\bibinfo{year}{1992}).

\bibitem[{\citenamefont{Jungen and Atabek}(1977)}]{Jungen_1977}
\bibinfo{author}{\bibfnamefont{C.}~\bibnamefont{Jungen}} \bibnamefont{and}
  \bibinfo{author}{\bibfnamefont{O.}~\bibnamefont{Atabek}},
  \bibinfo{journal}{J.\ Chem.\ Phys.} \textbf{\bibinfo{volume}{66}},
  \bibinfo{pages}{5584} (\bibinfo{year}{1977}).

\bibitem[{\citenamefont{Glass-Maujean et~al.}(2011)\citenamefont{Glass-Maujean,
  Jungen, Roudjane, and Tchang-Brillet}}]{Glass_2011}
\bibinfo{author}{\bibfnamefont{M.}~\bibnamefont{Glass-Maujean}},
  \bibinfo{author}{\bibfnamefont{C.}~\bibnamefont{Jungen}},
  \bibinfo{author}{\bibfnamefont{M.}~\bibnamefont{Roudjane}}, \bibnamefont{and}
  \bibinfo{author}{\bibfnamefont{W.-{\"U}.~L.} \bibnamefont{Tchang-Brillet}},
  \bibinfo{journal}{J.\ Chem.\ Phys.} \textbf{\bibinfo{volume}{134}},
  \bibinfo{pages}{204305} (\bibinfo{year}{2011}).

\end{thebibliography}

\end{document}


\newcommand{\cm}{cm$^{-1}$}
\newcommand{\B}{$B\,^1\Sigma^+_{\mathrm{u}}$}
\newcommand{\Bp}{$B'\,^1\Sigma^+_{\mathrm{u}}$}
\newcommand{\Bpp}{$B''\,^1\Sigma^+_{\mathrm{u}}$}
\newcommand{\C}{$C\,^1\Pi_{\mathrm{u}}$}
\newcommand{\Cplus}{$C\,^1\Pi_{\mathrm{u}}^+$}
\newcommand{\Cminus}{$C\,^1\Pi_{\mathrm{u}}^-$}
\newcommand{\D}{$D\,^1\Pi_{\mathrm{u}}$}
\newcommand{\Dplus}{$D\,^1\Pi_{\mathrm{u}}^+$}
\newcommand{\Dminus}{$D\,^1\Pi_{\mathrm{u}}^-$}
\newcommand{\Dp}{$D'\,^1\Pi_{\mathrm{u}}$}
\newcommand{\Dpp}{$D''\,^1\Pi_{\mathrm{u}}$}
\newcommand{\X}{$X\,^1\Sigma^+_{\mathrm{g}}$}
\newcommand{\dash}{\multicolumn{1}{c}{--}}

\maketitle

\begin{squeezetable}
\begin{table*}
\caption
{\label{table_freq_list_B}
  Observed transition frequencies of the \B--\X\ bands of D$_2$.
  The values are in \cm\ and in superscript the estimated uncertainty is given.
  Blended lines are indicated with $^b$.
}
\begin{ruledtabular}
\begin{tabular}{rdddddddd}
$J$ & P(J)        & R(J)        & P(J)        & R(J)        & P(J)        & R(J)        & P(J)        & R(J)        \\
\hline 
    & \multicolumn{2}{c}{0--0}  & \multicolumn{2}{c}{1--0}  & \multicolumn{2}{c}{2--0}  & \multicolumn{2}{c}{3--0}  \\
 0  &             &  90653.18^3 &             &  91594.78^3 &             &  92517.03^3 &             &  93421.10^3 \\
 1  &  90573.69^3 &  90632.74^3 &  91516.00^3 &  91572.90^3 &  92438.88^3 &  92493.90^3 &  93343.52^3 &  93396.84^3 \\
 2  &  90474.11^3 &  90572.25^3 &  91415.72^3 &  91510.26^3 &  92337.96^3 &  92429.39^3 &  93242.03^3 &  93330.64^3 \\
 3  &             &  90471.83^6 &  91275.39^3 &  91407.10^3 &  92196.36^3 &  92323.76^3 &  93099.31^3 &  93222.83^3 \\
 4  &             &             &  91095.59^6 &  91263.90^6 &  92014.73^6 &  92177.62^3 &  92915.99^3 &  93073.93^3 \\
\\
    & \multicolumn{2}{c}{4--0}  & \multicolumn{2}{c}{5--0}  & \multicolumn{2}{c}{6--0}  & \multicolumn{2}{c}{7--0}  \\
 0  &             &  94307.66^3 &             &  95177.16^3 &             &  96029.88^3 &             &  96866.12^b \\
 1  &  94230.61^3 &  94282.37^3 &  95100.59^3 &  95150.90^3 &  95953.76^3 &  96002.72^3 &  96790.41^3 &  96838.11^3 \\
 2  &  94128.60^3 &  94214.63^3 &  94998.10^3 &  95081.76^3 &  95850.81^3 &  95932.22^3 &  96687.04^3 &  96766.33^3 \\
 3  &  93984.82^3 &  94104.81^3 &  94853.37^3 &  94970.03^3 &  95705.20^3 &  95818.73^3 &  96540.56^3 &  96651.15^3 \\
 4  &  93799.97^3 &  93953.39^3 &  94667.09^3 &  94816.31^3 &  95517.60^3 &  95662.80^3 &  96351.70^3 &  96493.13^3 \\
\\
    & \multicolumn{2}{c}{8--0}  & \multicolumn{2}{c}{9--0}  & \multicolumn{2}{c}{10--0} & \multicolumn{2}{c}{11--0} \\
 0  &             &  97685.98^3 &             &  98489.67^3 &             &  99277.32^3 &             & 100049.18^3 \\
 1  &  97610.68^3 &  97657.15^3 &  98414.76^3 &  98460.06^3 &  99202.81^3 &  99246.95^3 &  99975.02^3 & 100018.10^3 \\
 2  &  97506.90^3 &  97584.17^3 &  98310.59^3 &  98385.90^3 &  99098.27^3 &  99171.69^3 &  99870.12^3 &  99941.76^3 \\
 3  &  97359.61^3 &  97467.36^3 &  98162.51^3 &  98267.57^3 &  98949.40^3 &  99051.83^3 &  99720.58^3 &  99820.53^3 \\
 4  &  97169.51^3 &  97307.37^3 &  97971.24^3 &  98105.67^3 &  98757.01^3 &  98888.12^3 &  99527.09^3 &  99655.05^3 \\
 5  &             &             &             &             &             &  98681.25^6               &             \\
\\
    & \multicolumn{2}{c}{12--0} & \multicolumn{2}{c}{13--0} & \multicolumn{2}{c}{14--0} & \multicolumn{2}{c}{15--0} \\
 0  &             & 100805.26^3 &             & 101546.07^3 &             & 102271.41^3 &             & 102981.69^3 \\
 1  & 100731.56^3 & 100773.54^3 & 101472.59^3 & 101513.64^3 & 102198.30^3 & 102238.27^3 & 102908.85^3 & 102948.01^3 \\
 2  & 100626.28^3 & 100696.12^3 & 101366.99^3 & 101435.26^3 & 102092.34^3 & 102158.84^3 & 102802.62^3 & 102867.83^3 \\
 3  & 100475.99^3 & 100573.46^3 & 101216.09^3 & 101311.38^3 & 101940.72^3 & 102033.56^3 & 102650.47^3 & 102741.61^3 \\
 4  & 100281.47^3 & 100406.22^3 & 101020.65^3 & 101142.69^3 & 101744.20^3 & 101863.06^3 & 102453.18^6 & 102570.52^3 \\
 5  &             & 100195.24^6               &             &             &             &             & 102342.13^6 \\
\\
    & \multicolumn{2}{c}{16--0} & \multicolumn{2}{c}{17--0} & \multicolumn{2}{c}{18--0} & \multicolumn{2}{c}{19--0} \\
 0  &             & 103676.91^3 &             & 104357.52^3 &             & 105023.31^3 &             &             \\
 1  & 103604.44^3 & 103642.53^3 & 104285.20^3 & 104322.80^3 & 104951.40^3 & 104987.75^3 & 105603.14^3 & 105636.08^3 \\
 2  & 103497.83^3 & 103561.23^3 & 104178.45^3 & 104241.31^3 & 104844.24^3 & 104904.70^3 & 105497.31^3 & 105553.10^3 \\
 3  & 103344.93^6 & 103433.46^3 & 104024.82^b & 104118.16^3 & 104690.24^6 & 104774.59^3 &             & 105422.16^3 \\
 4  & 103146.56^3 & 103259.86^3 & 103826.75^3 & 103933.12^3 & 104490.06^6 &             &             & 105244.39^3 \\
\\
    & \multicolumn{2}{c}{20--0} & \multicolumn{2}{c}{21--0} & \multicolumn{2}{c}{22--0} & \multicolumn{2}{c}{23--0} \\
 0  &             & 106312.08^3 &             & 106935.16^3 &             & 107544.74^3 &             & 108140.33^3 \\
 1  & 106240.76^3 & 106275.42^3 & 106864.24^3 & 106897.68^3 & 107473.90^3 & 107507.03^3 & 108069.90^3 & 108101.91^3 \\
 2  & 106133.05^3 & 106190.73^3 & 106756.07^3 & 106811.92^3 & 107365.70^3 & 107420.82^3 & 107961.26^3 & 108014.65^3 \\
 3  &             & 106058.45^3 & 106600.14^3 & 106678.36^b & 107209.41^6 & 107286.52^3 & 107804.37^3 & 107879.03^3 \\
 4  &             &             &             & 106497.64^6 &             &               107599.95^6 & 107695.73^3 \\
\\
    & \multicolumn{2}{c}{24--0} & \multicolumn{2}{c}{25--0} & \multicolumn{2}{c}{26--0} & \multicolumn{2}{c}{27--0} \\
 0  &             & 108722.69^3 &             & 109291.46^3 &             & 109847.39^3 &             & 110389.92^3 \\ 
 1  & 108652.38^3 & 108684.02^3 & 109221.50^3 & 109252.10^3 & 109777.46^3 & 109807.90^3 & 110320.40^3 & 110349.60^3 \\ 
 2  & 108543.66^3 & 108596.41^3 & 109112.41^3 & 109163.39^3 & 109668.30^3 & 109719.21^3 & 110210.86^3 & 110259.50^3 \\ 
 3  & 108386.49^6 & 108460.45^6 &             & 109025.79^3 & 109510.34^6 &             & 110052.26^6 &             \\ 
 4  & 108181.79^6 &             & 108748.72^6 & 108840.10^6               &             &             &             \\
\\
    & \multicolumn{2}{c}{28--0} & \multicolumn{2}{c}{29--0} & \multicolumn{2}{c}{30--0} & \multicolumn{2}{c}{31--0} \\
 0  &             & 110920.19^3 &             & 111436.88^3 &             & 111940.80^3 &             & 112433.32^3 \\
 1  & 110850.47^3 & 110881.05^3 & 111367.81^3 & 111395.68^b & 111872.50^3 & 111898.61^3 & 112364.65^3 & 112391.24^3 \\
 2  & 110741.15^3 & 110776.25^3 & 111257.80^3 & 111304.21^3 & 111761.75^3 & 111806.12^3 & 112254.22^3 & 112298.50^3 \\
 3  & 110583.56^3 & 110643.40^3 & 111098.08^6 & 111162.94^6 &             & 111663.72^6 & 112093.66^6 & 112155.59^6 \\
 4  &             & 110455.44^6               &             &             &             &             &             \\
\\
    & \multicolumn{2}{c}{32--0} & \multicolumn{2}{c}{33--0} & \multicolumn{2}{c}{34--0} & \multicolumn{2}{c}{35--0} \\
 0  &             & 112912.58^3 &             & 113379.86^3 &             & 113834.40^3 &             & 114276.76^3 \\
 1  & 112844.37^3 & 112869.76^3 & 113311.76^b & 113336.98^3 & 113766.50^3 & 113790.75^3 & 114208.96^3 & 114233.10^3 \\
 2  & 112733.52^3 & 112775.96^3 & 113200.80^3 & 113243.00^3 & 113655.30^3 & 113695.73^3 & 114097.70^3 & 114138.02^3 \\
 3  &             & 112631.73^6 &             & 113098.60^6               &             &             &             \\
\\
    & \multicolumn{2}{c}{36--0} & \multicolumn{2}{c}{37--0} & \multicolumn{2}{c}{38--0} & \multicolumn{2}{c}{39--0} \\
 0  &             & 114706.32^3 &             & 115123.58^3 &             & 115527.52^3 &             & 115918.74^3 \\
 1  & 114638.86^3 & 114661.88^3 & 115056.13^3 & 115079.18^3 & 115460.49^3 & 115482.22^3 & 115851.62^3 & 115873.92^3 \\
 2  & 114527.58^b & 114565.58^3 & 114944.50^3 & 114983.31^3 & 115348.44^3 & 115384.61^3 & 115739.67^3 & 115788.17^b \\
 3  &             & 114418.01^6 &             &             &             &             & 115576.46^6 & 115627.47^b \\
\\
    & \multicolumn{2}{c}{40--0} & \multicolumn{2}{c}{41--0} & \multicolumn{2}{c}{42--0} & \multicolumn{2}{c}{43--0} \\
 0  &             & 116295.64^3 &             & 116659.07^3 &             & 117006.06^3 &             & 117338.29^3 \\
 1  & 116229.07^3 & 116249.40^3 & 116592.16^3 & 116618.38^b & 116940.00^3 & 116958.81^3 & 117271.44^3 & 117283.73^3 \\
 2  & 116116.55^3 & 116150.42^3 & 116480.00^3 & 116506.22^3 & 116826.98^3 & 116858.28^3 & 117159.23^3 & 117184.84^3 \\
 3  &             & 115999.12^6 & 116320.99^b & 116356.13^b &             &             &             & 117030.91^6 \\
\\
    & \multicolumn{2}{c}{44--0} & \multicolumn{2}{c}{45--0} & \multicolumn{2}{c}{46--0} & \multicolumn{2}{c}{47--0} \\
 0  &             & 117650.23^3 &             & 117943.79^3 &             & 118212.04^3 &             & 118453.82^3 \\
 1  & 117584.80^3 & 117601.79^3 & 117877.81^3 & 117888.42^3 & 118147.39^3 & 118162.43^b & 118389.16^3 & 118405.55^3 \\
 2  & 117471.17^3 & 117499.40^3 & 117764.74^3 & 117787.38^3 & 118032.96^3 & 118057.22^3 & 118274.74^3 & 118293.17^3 \\
\\
    & \multicolumn{2}{c}{48--0} & \multicolumn{2}{c}{49--0} & \multicolumn{2}{c}{50--0} & \multicolumn{2}{c}{51--0} \\
 0  &             & 118660.52^b &             & 118824.72^3 &             & 118936.51^3 &             & 118989.87^3 \\
 1  & 118597.08^3 & 118608.22^3 & 118762.65^3 & 118770.94^b & 118874.79^6 & 118880.51^3 & 118929.17^6 & 118931.96^b \\
 2  & 118481.48^3 & 118500.00^3 &             & 118660.52^b &             & 118766.75^6 &             &             \\
\end{tabular}
\end{ruledtabular}
\end{table*}
\end{squeezetable}

\begin{squeezetable}
\begin{table*}
\caption
{\label{table_freq_list_Bp}
  Observed transition frequencies of the \Bp--\X\ bands of D$_2$.
  The values are in \cm\ and in superscript the estimated uncertainty is given.
  Blended lines are indicated with $^b$.
}
\begin{ruledtabular}
\begin{tabular}{rdddddddd}
$J$ & P(J)        & R(J)        & P(J)        & R(J)        & P(J)        & R(J)        & P(J)        & R(J)        \\
\hline 
    & \multicolumn{2}{c}{0--0}  & \multicolumn{2}{c}{1--0}  & \multicolumn{2}{c}{2--0}  & \multicolumn{2}{c}{3--0}  \\
 0  &             & 110841.93^3 &             & 112205.74^3 &             & 113490.86^3 &             & 114690.73^3 \\
 1  &             & 110834.61^3 & 112121.09^3 & 112195.62^3 & 113407.26^3 & 113478.67^3 & 114609.61^3 & 114673.86^3 \\
 2  & 110662.79^3 & 110793.64^3 & 112026.67^3 & 112150.61^3 & 113311.76^b & 113430.64^3 & 114511.68^3 & 114619.28^3 \\
 3  &             &             &             & 112070.98^3 & 113181.14^3 & 113347.12^3 & 114376.30^3 & 114527.58^b \\
 4  &             &             &             & 111957.17^6 &             &             &             & 114399.27^3 \\
\\
    & \multicolumn{2}{c}{4--0}  & \multicolumn{2}{c}{5--0}  & \multicolumn{2}{c}{6--0}  & \multicolumn{2}{c}{7--0}  \\
 0  &             & 115800.42^3 &             & 116803.71^3 &             & 117675.40^3 &             & 118371.08^3 \\
 1  & 115720.04^3 & 115781.77^3 & 116724.70^3 & 116782.31^3 & 117599.50^3 & 117648.04^3 & 118297.53^3 & 118338.63^3 \\
 2  & 115621.35^3 & 115724.02^3 & 116624.63^3 & 116720.42^3 & 117496.32^3 & 117577.68^3 & 118192.01^3 & 118259.91^3 \\
 3  & 115484.23^3 & 115627.47^b & 116484.76^3 & 116618.38^b & 117350.49^3 & 117464.67^3 & 118041.11^3 & 118134.87^3 \\
 4  &             & 115492.65^b & 116305.77^3 & 116476.80^3 & 117162.98^6 & 117309.45^3 & 117845.28^6 & 117963.59^6 \\
\\
    & \multicolumn{2}{c}{8--0}  & \multicolumn{2}{c}{9--0}  & \multicolumn{2}{c}{10--0} & \multicolumn{2}{c}{11--0} \\
 0  &             & 118761.30^3 &             & 118842.95^3 &             & 118916.30^3 &             & 118968.98^3 \\
 1  & 118695.04^3 & 118713.99^b & 118778.74^3 & 118792.43^3 & 118853.33^3 & 118862.83^3 & 118906.88^3 & 118913.78^3 \\
 2  & 118582.23^3 & 118611.96^3 & 118663.87^3 & 118684.60^3 & 118737.22^3 & 118752.93^3 & 118789.90^3 & 118801.23^3 \\
 3  & 118416.38^6 &             & 118494.91^6 & 118523.74^3 &             & 118586.92^6 &             &             \\
\\
    & \multicolumn{2}{c}{12--0} & \multicolumn{2}{c}{13--0} \\
 0  &             & 119005.19^3 &             & 119028.13^3 \\
 1  & 118943.69^3 & 118948.71^3 & 118967.48^3 & 118969.84^3 \\
 2  & 118826.11^3 & 118833.90^3 & 118849.08^3 &             \\
\end{tabular}
\end{ruledtabular}
\end{table*}
\end{squeezetable}

\begin{squeezetable}
\begin{table*}
\caption
{\label{table_freq_list_C}
  Observed transition frequencies of the \C--\X\ bands of D$_2$.
  The values are in \cm\ and in superscript the estimated uncertainty is given.
  Blended lines are indicated with $^b$.
}
\begin{ruledtabular}
\begin{tabular}{rddddddddd}
$J$ & P(J)        & Q(J)        & R(J)        & P(J)        & Q(J)        & R(J)        & P(J)        & Q(J)        & R(J)        \\
\hline 
    & \multicolumn{3}{c}{0--0}                & \multicolumn{3}{c}{1--0}                & \multicolumn{3}{c}{2--0}                \\
 0  &             &             &  99424.97^3 &             &             & 101085.37^b &             &             & 102677.96^3 \\
 1  &             &  99364.90^3 &  99427.14^3 &             & 101025.27^3 & 101085.37^b &             & 102617.85^3 & 102675.80^3 \\
 2  &  99245.89^3 &  99307.05^3 &  99400.50^3 & 100906.38^b & 100965.18^3 & 101055.48^3 & 102498.88^3 & 102555.55^3 & 102642.55^3 \\
 3  &  99129.63^3 &  99220.64^3 &  99345.17^3 & 100787.88^b & 100875.41^3 & 100995.74^3 & 102378.25^3 & 102462.48^3 & 102578.28^3 \\
 4  &  98985.85^3 &  99106.07^3 &  99261.43^3 & 100640.82^3 & 100756.44^3 & 100906.38^b & 102227.92^3 & 102339.11^3 & 102482.84^3 \\
 5  &             &             &             &             & 100608.74^6 & 100787.88^b &             & 102186.10^3 &             \\
\\
    & \multicolumn{3}{c}{3--0}                & \multicolumn{3}{c}{4--0}                & \multicolumn{3}{c}{5--0}                \\
 0  &             &             & 104203.85^3 &             &             & 105662.54^3 &             &             & 107059.40^3 \\
 1  &             & 104143.77^3 & 104199.30^3 &             & 105603.94^3 & 105660.43^3 &             & 106999.06^3 & 107051.09^3 \\
 2  & 104024.82^b & 104079.27^3 & 104162.25^3 & 105483.49^3 & 105537.34^3 & 105619.88^3 & 106880.30^3 & 106931.79^b & 107008.52^3 \\
 3  & 103901.77^3 & 103982.97^3 & 104088.99^3 & 105362.82^3 & 105437.82^3 & 105547.12^3 & 106753.52^3 & 106827.61^3 & 106931.79^b \\
 4  & 103747.61^3 & 103855.29^3 & 103998.54^3 & 105205.23^3 & 105305.87^3 & 105441.77^3 & 106593.88^3 & 106691.41^3 & 106821.20^3 \\
 5  &             &             &             &             & 105142.02^6 &             &             & 106522.37^3 & 106678.36^b \\
\\
    & \multicolumn{3}{c}{6--0}                & \multicolumn{3}{c}{7--0}                & \multicolumn{3}{c}{8--0}                \\
 0  &             &             & 108389.69^3 &             &             & 109655.36^3 &             &             & 110855.96^3 \\
 1  &             & 108329.47^3 & 108379.08^3 &             & 109595.27^3 & 109642.39^3 &             & 110796.21^3 & 110839.51^3 \\
 2  & 108210.59^3 & 108258.63^3 & 108333.13^3 & 109476.28^3 & 109522.28^3 & 109592.74^3 & 110676.89^3 & 110721.08^3 & 110801.26^3 \\
 3  & 108081.53^3 & 108152.72^3 & 108251.86^3 & 109344.87^3 & 109413.22^3 & 109505.61^3 & 110541.94^3 & 110608.83^3 & 110704.30^3 \\
 4  & 107918.49^3 & 108012.30^3 & 108135.19^3 & 109178.09^3 & 109268.61^3 & 109371.58^3 & 110386.60^6 & 110459.95^3 & 110576.50^3 \\
 5  &             & 107838.06^3 &             &             & 109089.06^6 &             &             &             &             \\
\\
    & \multicolumn{3}{c}{9--0}                & \multicolumn{3}{c}{10--0}               & \multicolumn{3}{c}{11--0}               \\
 0  &             &             & 111992.30^3 &             &             & 113060.71^3 &             &             & 114061.24^3 \\
 1  &             & 111931.61^3 & 111975.68^3 &             & 113000.46^3 & 113041.56^3 &             & 114001.13^3 & 114039.57^3 \\
 2  & 111813.20^3 & 111854.34^3 & 111920.38^3 & 112881.66^3 & 112920.98^3 & 112982.71^3 & 113882.18^3 & 113919.39^3 & 113976.97^3 \\
 3  & 111678.18^3 & 111738.85^3 & 111826.59^3 & 112744.09^6 & 112802.20^3 & 112884.22^3 & 113742.04^6 & 113797.21^3 & 113873.24^3 \\
 4  & 111505.76^6 & 111585.67^3 & 111694.64^3 &             & 112644.67^3 &             &             &             &             \\
 5  &             & 111395.68^b &             &             &             &             &             &             &             \\
\\
    & \multicolumn{3}{c}{12--0}               & \multicolumn{3}{c}{13--0}               & \multicolumn{3}{c}{14--0}               \\
 0  &             &             & 114991.35^3 &             &             & 115847.97^3 &             &             & 116626.95^3 \\
 1  &             & 114931.37^3 & 114967.02^3 &             & 115788.17^b & 115820.62^3 &             & 116567.49^3 & 116595.07^3 \\
 2  & 114812.28^3 & 114847.28^3 & 114900.24^3 & 115668.89^3 & 115701.61^3 & 115747.64^3 & 116447.88^3 & 116478.28^3 & 116530.47^3 \\
 3  & 114669.45^6 & 114721.50^3 & 114789.43^3 & 115523.05^6 & 115572.17^3 &             &             & 116344.91^3 & 116410.00^6 \\
\\
    & \multicolumn{3}{c}{15--0}               & \multicolumn{3}{c}{16--0}               & \multicolumn{3}{c}{17--0}               \\
 0  &             &             & 117322.91^3 &             &             & 117929.64^3 &             &             & 118437.02^3 \\
 1  &             & 117263.98^3 & 117297.52^3 &             & 117870.57^3 & 117900.48^3 &             & 118377.56^3 & 118395.12^3 \\
 2  & 117143.82^3 & 117171.92^3 & 117216.86^3 & 117750.54^b & 117775.32^3 & 117814.99^3 & 118257.93^3 & 118278.69^3 & 118312.61^3 \\
 3  &             & 117034.22^3 & 117092.23^6 &             & 117632.91^6 & 117684.02^6 &             & 118130.83^6 &             \\
 4  &             & 116851.28^3 &             &             &             &             &             &             &             \\
\\
    & \multicolumn{3}{c}{18--0}               & \multicolumn{3}{c}{19--0}               & \multicolumn{3}{c}{20--0}               \\
 0  &             &             & 118831.41^3 &             &             & 119085.89^3 &             &             &             \\
 1  &             & 118770.94^b & 118787.30^3 &             & 119026.04^3 & 119035.99^3 &             & 119097.21^6 &             \\
 2  & 118652.35^3 & 118667.84^3 & 118693.43^6 & 118906.88^3 &             & 118931.96^b &             &             &             \\
 3  &             & 118513.44^6 &             &             &             &             &             &             &             \\
\end{tabular}
\end{ruledtabular}
\end{table*}
\end{squeezetable}

\begin{squeezetable}
\begin{table*}
\caption
{\label{table_freq_list_D}
  Observed transition frequencies of the \D--\X\ bands of D$_2$.
  The values are in \cm\ and in superscript the estimated uncertainty is given.
  Blended lines are indicated with $^b$.
}
\begin{ruledtabular}
\begin{tabular}{rddddddddd}
$J$ & P(J)        & Q(J)        & R(J)        & P(J)        & Q(J)        & R(J)        & P(J)        & Q(J)        & R(J)        \\
\hline 
    & \multicolumn{3}{c}{0--0}                & \multicolumn{3}{c}{1--0}                & \multicolumn{3}{c}{2--0}                \\
 0  &             &             & 113222.93^3 &             &             & 114825.05^3 &             &             & 116359.52^3 \\
 1  &             & 113162.64^3 & 113223.62^3 &             & 114763.65^3 & 114825.35^3 &             & 116299.04^3 & 116356.13^b \\
 2  & 113043.89^3 & 113102.85^3 & 113194.62^3 & 114646.00^3 & 114701.66^3 & 114795.50^3 & 116180.43^3 & 116234.88^3 & 116320.99^b \\
 3  & 112926.05^6 & 113013.54^3 & 113135.86^6 & 114527.58^b & 114609.05^3 & 114734.97^3 & 116058.59^3 & 116139.03^3 & 116254.02^3 \\
 4  &             & 112895.16^6 &             & 114380.80^6 &             & 114645.53^3 & 115906.31^6 & 116012.02^3 & 116155.25^6 \\
\\
    & \multicolumn{3}{c}{3--0}                \\
 0  &             &             & 117831.41^3 \\
 1  &             & 117770.25^3 & 117827.01^3 \\
 2  & 117652.34^3 & 117703.98^3 & 117789.82^3 \\
 3  & 117529.47^3 & 117604.96^3 & 117719.59^3 \\
 4  & 117375.17^3 & 117473.72^3 & 117616.27^3 \\
\end{tabular}
\end{ruledtabular}
\end{table*}
\end{squeezetable}